\documentclass[12pt]{elsart}
\usepackage{amstext,amsfonts,amssymb}
\def\ope[#1][#2]{\mathord{#2\over{\ifnum#1=1 {z-w} \else {(z-w)^{#1}}\fi}}}
\def\sope[#1/#2][#3]{\mathord{#3\over{\ifnum#2=1{\ifnum#1=1 {Z-W} \else 
{(Z-W)^{#1}}\fi} \else {(Z-W)^{#1/#2}}\fi}}}
\def\pair<#1,#2>{\mathop{\left\langle#1\mathbin{,}#2\right\rangle}\nolimits}
\def\fr#1#2{{\textstyle {#1\over#2}}}
\def\shalf{\mathord{\scriptscriptstyle 1\over \scriptscriptstyle 2}}
\newcommand{\ad}{\mathrm{ad}}
\renewcommand{\d}{\partial}
\newcommand{\reg}{\text{reg}}
\renewcommand{\gg}{{\ensuremath{\mathfrak g}}}
\newcommand{\gh}{{\ensuremath{\mathfrak h}}}
\newcommand{\gt}{{\ensuremath{\mathfrak t}}}
\newcommand{\gs}{{\ensuremath{\mathfrak s}}}
\newcommand{\sG}{{\ensuremath{\mathsf G}}}
\newcommand{\sT}{{\ensuremath{\mathsf T}}}
\newcommand{\sJ}{{\ensuremath{\mathsf J}}}
\newcommand{\1}{{\ensuremath{\mathbf 1}}}
\newcommand{\C}{{\ensuremath{\mathbb C}}}
\newcommand{\Z}{{\ensuremath{\mathbb Z}}}
\newcommand{\tA}{{\ensuremath{\mathbb A}}}
\newcommand{\tB}{{\ensuremath{\mathbb B}}}
\newcommand{\tC}{{\ensuremath{\mathbb C}}}
\newcommand{\tG}{{\ensuremath{\mathbb G}}}
\newcommand{\tH}{{\ensuremath{\mathbb H}}}
\newcommand{\tJ}{{\ensuremath{\mathbb J}}}
\newcommand{\tO}{{\ensuremath{\mathbb O}}}
\newcommand{\tT}{{\ensuremath{\mathbb T}}}
\newcommand{\tX}{{\ensuremath{\mathbb X}}}
\newcommand{\tY}{{\ensuremath{\mathbb Y}}}
\newcommand{\cA}{{\ensuremath{\mathcal A}}}
\newcommand{\bcD}{\bar\nabla_{\bar\cA}}
\newcommand{\cD}{\nabla_{\cA}}
\newcommand{\NPB}[3]{{\rm Nucl. Phys.} {\bf B#1} (#2) #3}
\newcommand{\CMP}[3]{{\rm Comm. Math. Phys.} {\bf #1} (#2) #3}
\newcommand{\PRD}[3]{{\rm Phys. Rev.} {\bf D#1} (#2) #3}
\newcommand{\PRL}[3]{{\rm Phys. Rev. Lett.} {\bf #1} (#2) #3}
\newcommand{\PLB}[3]{{\rm Phys. Lett.} {\bf #1B} (#2) #3}

\newcommand{\PTP}[3]{{\rm Prog. Theor. Phys.} {\bf B#1} (#2) #3}

\newcommand{\JETP}[3]{{\rm Sov. Phys. JETP} {\bf B#1} (#2) #3}
\begin{document}
\begin{frontmatter}
\title{Nonreductive WZW models and their CFTs, II: $N{=}1$ and $N{=}2$
cosets}
\author[QMW]{Jos\'e~M~Figueroa-O'Farrill\thanksref{emailjmf}\thanksref{EPSRC}}
and
\author[ICTP]{Sonia Stanciu\thanksref{emailss}\thanksref{IC}}
\address[QMW]{Department of Physics, Queen Mary and Westfield College,
Mile End Road, London E1 4NS, UK}
\address[ICTP]{ICTP, P.O. Box 586, I-34100 Trieste, ITALY}
\thanks[emailjmf]{\tt mailto:j.m.figueroa@qmw.ac.uk}
\thanks[emailss]{\tt mailto:sonia@ictp.trieste.it}
\thanks[EPSRC]{Supported in part by the EPSRC under contract
GR/K57824.}
\thanks[IC]{Address after October 1996: Theoretical Physics Group,
Blackett Laboratory, \\ Imperial College, London, UK}
\begin{abstract}
In \cite{FSnr} we started a programme devoted to the systematic
study of the conformal field theories derived from WZW models based on
nonreductive Lie groups.  In this, the second part, we continue this
programme with a look at the $N{=}1$ and $N{=}2$ superconformal field
theories which arise from both gauged and ungauged supersymmetric WZW
models.  We extend the supersymmetric (affine) Sugawara and coset
constructions, as well as the Kazama--Suzuki construction to general
self-dual Lie algebras.
\end{abstract}
\end{frontmatter}

\section{Introduction}

One of the most important problems in string theory is the
construction of exact (super)string backgrounds.  As a string
propagates in a manifold $M$, it interacts with it via the following
geometric data: a metric $g$, an antisymmetric two-form $b$, and a
scalar field $\varphi$.  Consistency imposes severe restrictions on
the {\sl background\/} $(M,g,b,\varphi)$, equivalent to demanding the
exact conformal invariance of the effective two-dimensional quantum
field theory on the string world-sheet.  For a general string
background $(M,g,b,\varphi)$, this translates into complicated coupled
nonlinear partial differential equations for $g$, $b$, and $\varphi$
involving an infinite number of perturbative $\alpha'$ corrections:
the vanishing of the exact $\beta$-function.  The exact form of the
equations is therefore not known, and although presumably starting
from any classical background one can flow under the renormalisation
group to a nearby exact background, it is a fact that very few exact
string backgrounds are known explicitly.

One way to construct exact string backgrounds is to start with an
exact string background and perform operations which preserve both the
conformal invariance and the spacetime interpretation of string
propagation.  For example, if we start with a string propagating on a
flat Minkowski spacetime, one can, via toroidal compactifications and
orbifold constructions, reach other more realistic exact string
backgrounds.  More generally, one can start with string propagating on
a group manifold.  For string propagation to be consistent, however,
the group must possess a bi-invariant metric; that is, a metric
invariant under both left and right multiplications. Equivalently its
Lie algebra must be self-dual (see, for example, \cite{FSsd}).  Every
compact Lie group has a self-dual Lie algebra, as can be easily shown
by averaging over the group using the Haar measure, but these are not
the only Lie groups with this property.  The conformal field theory
describing string propagation on a group manifold is the WZW model.
The conformal invariance of the WZW model derives, via the affine
Sugawara construction, from its huge semi-local symmetry.  Gauging
some of this symmetry one can then obtain other exact string
backgrounds, some of which possess a spacetime interpretation as
string propagation on coset spaces.

Until relatively recently, most of the work on the construction of
string backgrounds starting from WZW models (or their supersymmetric
generalisations) was concerned with compact (or more generally,
reductive) Lie groups.  Reductive Lie groups are (up to coverings)
direct product of semisimple and abelian Lie groups, hence this class
of string backgrounds also comprises the flat spacetimes and the
toroidal compactifications.  But ever since the work of Nappi \&
Witten \cite{NW}, in which an exact four-dimensional string background
describing a gravitational wave was obtained from the WZW model
corresponding to a nonsemisimple (in fact, solvable) Lie group, the
possibility has arisen of considering more general Lie groups.  In a
recent paper \cite{FSnr}, of which the present paper is a continuation and
will hereafter be referred to as Part I, we have analysed in detail
the construction, conformal invariance and gauging of WZW models based
on nonreductive Lie groups, as well as the conformal field theories
they give rise to.  We refer the reader to this paper for references
to the work on nonreductive WZW models.

In the present paper we extend the results of Part I to supersymmetric
WZW models and their $N{=}1$ and $N{=}2$ supersymmetric cosets.
Whereas the relation between gauged WZW models and coset constructions
was already well-understood in the reductive case (see Part I and
references therein), this happy state of affairs did not persist in
the supersymmetric situation.  Two coset constructions are known which
yield superconformal field theories: the one by Goddard, Kent \& Olive
(GKO) in \cite{GKO} and the one by Kazama \& Suzuki (KS) in
\cite{KazamaSuzuki} (see also \cite{Schweigert}).

The GKO construction is essentially a coset construction \cite{coset}
of the form $\gg\times\gg /\gg$ where the first $\gg$ in the numerator
is realised as a WZW model and the second $\gg$ as free fermions in
the adjoint representation; that is, the numerator corresponds to a
supersymmetric WZW model.  This theory is superconformal, but the
superconformal symmetry found by GKO after quotienting seems
accidental; that is, it is not related to the superconformal symmetry
of the unquotiented theory.  A path integral derivation of the GKO
construction, starting from a SWZW model, was given in
\cite{Schnitzer} (see also \cite{Nakatsu}).  This construction starts
by gauging a bosonic diagonal subalgebra $\gh$ in the SWZW model.  A
superconformal theory is recovered only when $\gh=\gg$, precisely as
in the GKO construction.  With hindsight this result is to be
expected: supersymmetry demands a delicate balance between the
fermionic and bosonic degrees of freedom, which is upset if one gauges
a symmetry which only gets rid of, say, bosonic degrees of freedom.
What is surprising about the results of \cite{GKO} and
\cite{Schnitzer} is that the full diagonal gauging should give rise to
a superconformal theory at all.

On the other hand, the KS construction is a natural superconformal
coset where the superconformal symmetry is preserved along the way.
However a satisfactory lagrangian description for the KS coset has
taken longer to appear.  In his work on topologically twisted KS
cosets, Witten \cite{Witten} wrote down without derivation a lagrangian
for the coset theory.  It was Tseytlin \cite{Tseytlin} (although see
earlier work by Nojiri \cite{Nojiri} in a particular case) who first wrote
down a path-integral derivation of the KS construction for an
arbitrary compact Lie group, starting from a SWZW model and gauging a
diagonal subgroup of both bosonic and fermionic symmetries.  Although
the construction in \cite{Tseytlin} assumes that the Lie group be compact,
a similar construction extends to the general case, as we will see
below in more detail.  These results notwithstanding, a conformal
field-theoretical derivation of the KS coset constructions from a
gauged WZW model, in the style of \cite{GWZW}, did not exist even in the
reductive case.

Even less is known in the nonreductive case.  Indeed, at the time of
writing the only work that had been done in this direction has been
the generalisation of the $N{=}1$ Sugawara construction \cite{Nouri}.  The
original motivation of this paper was to fill this gap.  Some of those
results in this paper which transcend the nonreductive programme
started in Part~I have already been announced in our letter \cite{FSn12},
where among other things we present for the first time (although see
also \cite{Rhedin}) a conformal field theoretical proof of the
superconformal invariance of the gauged SWZW model.

Although the SWZW model with or without gauging is always invariant
under an $N{=}1$ superconformal algebra, it sometimes admits an
extended $N{=}2$ superconformal symmetry.  For arbitrary
supersymmetric $\sigma$-models, Hull and Witten \cite{HullWitten} wrote
down the necessary conditions for the existence of a (linearly
realised) such $N{=}2$ supersymmetry.  When specialised to a SWZW
model, these conditions translate into algebraic conditions on the Lie
algebra, which were first obtained by the Leuven group \cite{Leuven}.
These conditions were later re-interpreted in terms of Manin triples
by Parkhomenko \cite{Parkhomenko}, and further elaborated by Getzler
\cite{GetzlerMT}, who also discussed the coset construction.  All these
papers treat only the reductive case, but many of their results extend
straightforwardly to the nonreductive case, as was shown by Mohammedi
\cite{NouriN=2} and also by one of us \cite{Fc=9N=2}.  In this latter
paper the connection with Manin triples was re-established and a
classification is given of those SWZW models based on a solvable Lie
group which admit an $N{=}2$ superconformal symmetry with central
charge $c{=}9$.

In the same way, under some conditions, the KS coset construction
admits an $N{=}2$ superconformal symmetry; and in fact, with
hindsight, the $N{=}2$ superconformal symmetry of the SWZW model (when
it exists) is a special case of the $N{=}2$ superconformal invariance
of the KS coset construction, where one gauges the trivial group.  In
this paper we will extend to the nonreductive case the $N{=}2$ KS
coset construction, and in particular show how the extra symmetry
comes induced from the gauged SWZW model.

We now come to a guided tour through the contents of this paper, but
first a word on superspace versus components.  Although in practice
many practitioners in this field feel they have to choose between one
convention or another, it is abundantly clear that each convention has
its virtues as well as its shortcomings.  As a result of this
tendency, much of the literature contains partial results which have
been obtained using one of the approaches, and the emerging
picture---even in the reductive case---is not completely
satisfying. Motivated by this, we will work both in superspace and in
components, in an attempt to clarify at the same time the various
connections between the two approaches in the context of the SWZW
model.

Maybe the most natural way of defining the $N{=}1$ SWZW model is to
covariantise the WZW model supersymmetrically.  Formal similarity
aside, the superspace formulation also turns out to yield the most
direct and non-ambiguous proof of the superconformal invariance of the
theory.  We begin thus in Section 2 by describing the $N{=}1$ SWZW
model based on a general self-dual Lie group in terms of superfields.
We write down the action, examine its symmetries, determine its
current algebra and prove its conformal invariance using the self-dual
$N{=}1$ Sugawara construction.

We turn then our attention to the gauged SWZW model.  In Section 3,
still working in superfields, we gauge a diagonal subgroup of the
isometries of the model.  Having done this, in Section 4 we prove the
superconformal invariance of the model and determine exactly the
corresponding SCFT.  More precisely, we define the self-dual KS coset
and we prove that this is precisely the superconformal field theory of
the gauged SWZW model.  This is done using the BRST formalism in
superspace.  The calculations rely on the superspace operator product
expansion, and in the Appendix we write down the ``superspace
Borcherds axioms'' for what we believe to be the first time (although
see \cite{Kris}).

The main appeal of the component formulation of the SWZW theory is its
apparent simplicity.  After eliminating the auxiliary fields, the SWZW
action can be written as a WZW model coupled axially to Weyl fermions
in the adjoint representation.  For the gauged SWZW model, as remarked
above, Witten \cite{Witten} wrote down an action which describes the
gauged WZW model coupled minimally to Weyl fermions on the coset
directions. Its simplicity notwithstanding, this formulation still
reserves us a few surprises.

%Indeed, as it turns out, there is more than one way of parametrising
%the basic superfield in terms of bosonic and fermionic fields; and
%while this ambiguity creates no problems at the classical level, since
%all these parametrisations yield equivalent SWZW actions, it certainly
%does create problems at the quantum level, where the different
%parametrisations give rise to different theories.  Also, to this
%ambiguity, we have to add the fact that it is not clear, from the path
%integral point of view, whether one should leave the fermions coupled
%to the bosonic field or rather if one can decouple them prior to
%gauging the model.  One of our aims is to bring some clarifications
%regarding these points.
%There are basically three levels at which we can perform the breaking
%into components: at the classical level (at the level of the action),
%before and after gauging it, and at the quantum level, in the final
%path integral.  The last one made the subject of our previous paper
%\cite{FSn12}, hence here we will only consider the first two.

An immediate choice which presents itself is whether to break up the
action into components before or after gauging.  The symmetries
involved are sufficiently different that the equivalence of the two
methods is not obvious.  Therefore we treat both cases and show that
they are equivalent and in fact equivalent to Witten's action.

In Section 5 we break the SWZW action into components.  As we will
see, the parametrisation of the superfield into bosonic and fermionic
components is not unique.  Different parametrisations are related by
chiral gauge transformations, which although perfectly valid in the
classical theory, are anomalous quantum mechanically.  This means that
different parametrisations yield different quantum theories.  As we
will see, the superconformal invariance of the quantum theory will
give us the key to solve the ambiguity in the parametrisation of the
superfield.  Once having chosen a parametrisation, we eliminate the
auxiliary fields, and determine its symmetries and the associated
conserved currents.

Section 6 is dedicated to the gauged SWZW model and to the derivation
of Witten's action.  We start by gauging the component action, which
as we will see, is not completely straightforward.  We then take the
gauged SWZW model in superfields and show that after eliminating the
auxiliary fields, the resulting action is precisely Witten's action;
and moreover that it agrees with he action obtained by gauging the
component SWZW model.

In Section 7, we consider the self-dual Kazama--Suzuki cosets which
possess an extended $N{=}2$ superconformal symmetry.  For the results
of this section it is necessary to work in components, since the
expression for the generators of the extended supersymmetry is not
local in the original superfields.  In this section we derive the
conditions under which a supersymmetric cosets admits $N{=}2$
supersymmetry and show that this symmetry is present already in the
gauged SWZW model description.

Let us close with a notational remark.  We will be using freely the
notation and results of \cite{FSnr} and any reference to an equation,
section, theorem in that paper will be prefixed by an ``I'', so that,
for example, equation (I.m.n) refers to equation in (m.n) in \cite{FSnr}.

\section{The $N{=}1$ SWZW model}

In this section we introduce the $N{=}1$ supersymmetric WZW model
associated with a Lie group possessing a bi-invariant metric.
Starting with the action written in terms of superfields, we derive
the classical and quantum algebras of currents, after which we review
the $N{=}1$ Sugawara construction.  This proves the superconformal
invariance of the supersymmetric WZW model.

\subsection{The $N{=}1$ SWZW model}

We shall construct the $N{=}1$ supersymmetric WZW model (SWZW model,
for short) as a supersymmetric covariantisation of the WZW model.  The
WZW model is manifestly a classical conformal field theory, and the
fact that this persists at the quantum level can be attributed, via
the Sugawara construction, to the affine Lie algebra of its conserved
currents.  Similarly, the SWZW model, when written in superfields, can
be seen to be manifestly a classical superconformal field theory; and
the $N{=}1$ Sugawara construction will guarantee that this continues
to hold at the quantum level.

The data defining this model is a connected Lie group $G$ possessing a
bi-invariant metric; that is, a metric invariant under both right and
left multiplication in $G$. This condition can also be understood in
terms of the Lie algebra $\gg$.  Any metric on $\gg$ can be
automatically promoted to a left-invariant metric on $G$ simply by
identifying the Lie algebra with the left-invariant vector fields.  A
necessary and sufficient condition for this metric to be also
right-invariant is that the metric on the Lie algebra be ad-invariant;
that is, invariant under the infinitesimal adjoint action.  Not every
Lie algebra will possess such a metric: those which do are called {\sl
self-dual}.  Semisimple and abelian Lie algebras comprise a small
subclass of the self-dual Lie algebras, which themselves for a very
small subclass of all Lie algebras.  Although the structure of
self-dual Lie algebras is mostly under control, a classification is
still lacking.  This is an interesting problem, for as shown in
\cite{Nouri} (see also \cite{FSsug}), self-dual Lie algebras are in
one-to-one correspondence with (S)WZW models.  For a recent survey of
results on self-dual Lie algebras see \cite{FSsd}.

In this section we present the superfield description of the SWZW
model.  Our conventions for superspace and superspace operator product
expansions are summarised in Appendix A, which the reader is invited
to peruse at this stage.  The fundamental fields will therefore be
superfields $\tG(Z,\bar Z)$, where $Z = (z,\theta)$ and $\bar Z = (\bar
z,\bar\theta)$ are coordinates in an $N{=}1$ super-Riemann surface
$\Sigma_S$ whose body is a Riemann surface $\Sigma$.  The
$\theta$-independent component of $\tG(Z,\bar Z)$ is then a $G$-valued
field $g(z,\bar z)$ on $\Sigma$.  Although the following nomenclature
is strictly speaking not correct, we will speak of $\tG$ as a
$G$-valued superfield.  We will have more to say about what $\tG$ is
later on in Section 5, when we discuss the component action.

Classically, the supersymmetric WZW model is defined by the following
action, directly generalising the one in \cite{SWZW} for the reductive
case:
\begin{equation}
I_\Omega[\tG] = 
\int_{\Sigma_S}\pair<\tG^{-1}D\tG,\tG^{-1}{\bar D}\tG> +
\int_{B_S}\pair<{\widetilde \tG}^{-1}\d_t{\widetilde \tG},
       \left[{\widetilde \tG}^{-1}D{\widetilde \tG},
       {\widetilde \tG}^{-1}\bar D{\widetilde \tG}\right]>~,\label{eq:SWZW}
\end{equation}
where $\widetilde \tG$ is an extension of $\tG$ to the cone over
$\Sigma_S$ -- a supermanifold $B_S$ with boundary $\Sigma_S$.  The
obstruction to this extension cancels due to the vanishing of
$\pi_2(G)$, which still holds for a nonreductive Lie group, since
any Lie group has the homotopy type of its maximal compact subgroup,
which is reductive.  The integrals in the above expression denote both
the geometric integral over $\Sigma$ and the Berezin integral over the
fermionic coordinates.  The subscript $\Omega$ keeps track of the
dependence of the action on the metric $\Omega_{ab} = \pair<X_a,X_b>$,
relative to a basis $\{X_a\}$ for $\gg$, which we fix once and for
all.  For a simple Lie group, all metrics are proportional, so one
usually fixes a metric and represents the dependence on the metric by
a parameter multiplying the action.  Upon quantisation, this parameter
usually becomes the level of the SWZW model.  On the other hand, a
general self-dual Lie group will have more than one metric, hence the
need to keep track of it.  As in the WZW model, we will assume that
the metric is nondegenerate, for otherwise not all fields would be
dynamical; that is, the theory would be constrained.  This does not
represent any loss of generality, since eliminating the non-dynamical
fields, which take values in an invariant subgroup, yields a SWZW
model on the quotient group which does inherit a nondegenerate metric.

The relative coefficient in (\ref{eq:SWZW}) has been chosen for
reasons which are standard \cite{WZW} but which we will review below.
The quantum field theory will be described by the path integral
\begin{displaymath}
Z = \int [d\tG] e^{-I_\Omega[\tG]}~.
\end{displaymath}
Independence of the quantum theory on the extension $\widetilde \tG$
will in general choose a discrete set of possible metrics (in the case
of a simple Lie group, this statement is simply the quantisation of
the level); although for some nonsemisimple Lie groups (e.g., if $G$
is solvable), this will not be the case.

Although it may not be obvious at this point, the action $I_\Omega[\tG]$
is invariant under any automorphism of the self-dual Lie group; that
is, any transformation which preserves both the Lie brackets {\em
and\/} the metric.  For $G$ a simple Lie group, this group is
essentially $G\times G$, corresponding to left and right
multiplications; but for a general self-dual Lie group, the full
automorphism group may be bigger.  Nevertheless we will focus in this
paper only on $G\times G$.  The proof that the action
$I_\Omega[\tG]$ in invariant under $G\times G$ will be delayed until
Section 5 when we discuss the component action.

In fact, thanks to the choice of relative coefficients in the action,
$I_\Omega[\tG]$ enjoys an infinite-dimensional ``semi-local'' symmetry:
$G(Z) \times G(\bar Z)$:
\begin{equation}
\tG(Z,\bar Z) \mapsto \Omega^{-1}(Z)\tG(Z,\bar Z){\bar\Omega}(\bar
Z)~,\label{eq:swzwinv}
\end{equation}
where $\Omega$ (resp.~$\bar\Omega$) is a chiral (resp.~antichiral)
superfield: $\bar D\Omega = 0$ and $D\bar \Omega = 0$.  This condition
will in general reduce the number of components of the superfield, as
well as forcing the remaining component fields to depend
(anti)hol\-o\-mor\-phically on the bosonic coordinates $(z,\bar z)$ (see
Section 5).

This invariance gives rise to the following conserved currents
\begin{equation}
\tJ(Z) = - D\tG\tG^{-1},\qquad
\bar\tJ(\bar Z) = \tG^{-1}{\bar D}\tG~,\label{eq:curr}
\end{equation}
satisfying the conditions ${\bar D}\tJ =D{\bar\tJ}=0$.  If we take
$\tJ$ and $\bar\tJ$ as the dynamical variables of the SWZW model and
treat the conservation laws as the equations of motion, then we obtain
for the fundamental Poisson brackets 
\begin{displaymath}
\left\{ \tJ_a (Z),\tJ_b (W)\right\} = \left(\Omega_{ab}D_W
+ {f_{ab}}^c \tJ_c(W)\right)\delta (Z-W)~,
\end{displaymath}
where $Z=(z,\theta)$, $W=(w,\varphi)$ and $\delta(Z-W) = \delta(z-w)
(\theta- \varphi)$.  Upon quantisation, the above Poisson brackets
yield the current algebra encoded in the following supersymmetric
operator product expansion:
\begin{equation}
\tJ_a (Z)\tJ_b (W) = \sope[1/1][\Omega_{ab}] +
 \sope[1/2][{f_{ab}}^c \tJ_c (W)] + \reg~.\label{eq:sope}
\end{equation}
(Notice that the above notation is ambiguous since the fields do not
always commute with the half-integral powers of the superinterval.  We
will follow the convention that even though we write the fields on top
of the superintervals, they are understood to appear to the right.)
Similar formulas hold for $\bar{\tJ}$.  In other words, these currents
satisfy an $N{=}1$ affine algebra $\widehat\gg_{N{=}1}$
\cite{KacTodorov}, whose central extension is defined by the metric in
the SWZW model.  This huge symmetry underpins the exact superconformal
invariance of the SWZW model.  The proof of this fact relies on the
$N{=}1$ Sugawara construction, to which we now turn.

\subsection{The $N{=}1$ Sugawara construction}

We start with the current algebra (\ref{eq:sope}).  By an $N{=}1$
Sugawara construction we mean the construction of an $N{=}1$
superconformal algebra out of (normal ordered) products in the
supercurrents $\tJ_a(Z)$, with the property that the supercurrents are
primary superfields of weight $\half$.  We take therefore as a general
Ansatz for the energy-momentum tensor
\begin{equation}
\tT(Z) = A^{ab}(D\tJ_a\tJ_b)(Z) + B^{abc}(\tJ_a (\tJ_b \tJ_c))(Z) +
          C^a\d \tJ_a (Z)~,\label{eq:sugt}
\end{equation}
with $A^{ab}$, $B^{abc}$, and $C^a$ yet unspecified coefficients.  If 
we now impose that the $\tJ_a(Z)$ be primary superfields of weight
$\half$, that is,
\begin{displaymath}
\tT(Z)\tJ_a(W) = \sope[3/2][\half\tJ_a(W)] +
\sope[1/1][\half D\tJ_a(W)] + \sope[1/2][\d\tJ_a(W)] + \reg~,
\end{displaymath}
we obtain that $C^a=0$, that $\Omega_{ab}$ must be invertible with
inverse $\Omega^{ab}$ and that the remaining coefficients are given by
\begin{displaymath}
A^{ab} = \half \Omega^{ab},\qquad B^{abc} = {\fr{1}{6}}f^{abc}~,
\end{displaymath}
where we use $\Omega^{ab}$ and $\Omega_{ab}$ to raise and lower
indices.  A straightforward calculation then shows that the
supersymmetric energy-momentum tensor (\ref{eq:sugt}) obeys the $N{=}1$
superconformal algebra with central charge
\begin{displaymath}
c_{\gg} \equiv {\fr{3}{2}}\dim\gg - {\fr{1}{2}}\Omega^{ab}\kappa_{ab}~,
\end{displaymath}
where $\kappa_{ab}$ stands for the Killing form of $\gg$, which for a
general self-dual Lie algebra need not be nondegenerate.

Summarising, we have seen that for the $N{=}1$ Sugawara construction
to exist it is necessary and sufficient that $\Omega_{ab}$, given by
(\ref{eq:sope}), be invertible, which coincides with the condition
required for the supersymmetric WZW model to define an unconstrained
field theory.  Notice that this is unlike the bosonic case, where we
had two different conditions that could only be simultaneously
satisfied because of the particular structure of self-dual Lie
algebras (see Theorem I.3.6 and also \cite{FSsd}).

\section{The gauged SWZW model}

After having seen that the SWZW model provides a lagrangian
realisation of the $N{=}1$ Sugawara construction, we now turn our
attention to the gauged SWZW model and to the superconformal field
theories (SCFT) that this procedure gives rise to.  We will consider
here the case of the diagonal gauging for which, by manipulating the
functional integral, we will be able to exhibit the resulting quantum
field theory as a SCFT whose energy-momentum tensor agrees with that
of an $N{=}1$ coset construction.

\subsection{Gauging the SWZW model}

We consider the problem of gauging a diagonal subgroup $H\subset G
\times G$. In other words we want to partially promote the semi-local
symmetry (\ref{eq:swzwinv}) restricted to a diagonal subgroup, to a
local invariance under transformations of the form:
\begin{equation}
\tG(Z,{\bar Z}) \mapsto \Lambda(Z,{\bar Z})^{-1}\,\tG(Z,{\bar Z})\,\Lambda
(Z,{\bar Z})~,\label{eq:gauge1}
\end{equation}
in the obvious notation.  For this we have to introduce gauge
superfields $\tA$ and $\bar\tA$, fermionic with weight $\half$, with
values in the complexified Lie algebra $\gh^{\C}$ of $H$, and
which transform under gauge transformations according to
\begin{eqnarray}
\tA &\mapsto& \Lambda^{-1}(D + \tA)\Lambda~,\nonumber\\
{\bar\tA} &\mapsto& \Lambda^{-1}(\bar D +
                                {\bar\tA})\Lambda~.\label{eq:gauge2}
\end{eqnarray}

Gauging the SWZW model means constructing an extension
$I_\Omega[\tG,\tA,\bar\tA]$ of (\ref{eq:SWZW}) which is invariant under
(\ref{eq:gauge1}) and (\ref{eq:gauge2}).  Using the Noether procedure we
obtain
\begin{equation}
I_\Omega[\tG,\tA,\bar\tA] = I_\Omega[\tG] - 2 \int_{\Sigma_S}
                          \pair<\tA,\bar\tJ> + \pair<\tJ,\bar\tA> -
                          \pair<\tA,\bar\tA> + 
                          \pair<\tA,\tG^{-1}{\bar\tA}\tG>.\label{eq:gaugact}
\end{equation}
Notice that, since the gauge superfields have no kinetic term, they
can be thought of as Lagrange multipliers: they introduce constraints
at the level of the classical theory, which consist in setting the 
$H$-current equal to zero.

The quantum theory is described by the path integral
\begin{displaymath}
Z = \int [d\tG][d\tA][d\bar\tA] e^{-I_\Omega[\tG,\tA,\bar\tA]}~.
\end{displaymath}
As discussed in Section I.4 for the nonsupersymmetric case, we
choose the holomorphic gauge $\bar\tA = 0$; in the absence of gauge
anomalies this will introduce in the gauge-fixed path integral a
Faddeev-Popov determinant, $\det\bar D$, which can be formally
expressed as the path integral
\begin{displaymath}
\det\bar D = \int [d\tB][d\tC] e^{-\int_{\Sigma_s}
\pair<\tB, \bar D \tC>}~,
\end{displaymath}
where $(\tB,\tC)$ are $N{=}1$ Faddeev-Popov ghost superfields.  $\tC$
is a fermionic weight zero $\gh^\C$-valued superfield, whereas $\tB$
is bosonic, has weight $\half$ and takes its values in the dual
$(\gh^\C)^*$.  The $\pair<-,->$ above indicates the dual pairing
between $\gh^\C$ and $(\gh^\C)^*$, together with superfield
multiplication.

The remaining gauge superfield $\tA$ can be parametrised by $\tA =
-D\tH\tH^{-1}$, where $\tH$ is an $H$-valued superfield.  We
now use the supersymmetric version of the Polyakov-Wiegmann identity
\cite{PW}, which holds for any self-dual Lie algebra, to express
$I_\Omega[\tG,\tA,0]$ in terms of the original SWZW action as follows:
\begin{displaymath}
I_\Omega[\tG,\tA,0] = I_\Omega[\tG\tH] - I_\Omega[\tH]~.
\end{displaymath}
At the quantum level, the change of variables from $\tA$ to $\tH$
modifies the functional measure of the path integral by a jacobian
factor, $\det D_{\tA}$, where $D_{\tA}$ denotes the holomorphic component
of the covariant derivative, $D_{\tA}=D+[\tA,-]$, acting on
$\gh^\C$-valued superfields.  We represent this determinant as:
\begin{displaymath}
\det D_{\tA} = \int [d\bar\tB][d\bar\tC] e^{-\int_{\Sigma_s}
\pair<\bar\tB,D_{\tA}\bar\tC>}\,
\end{displaymath}
where now $(\bar\tB, \bar\tC)$ are $(\gh^\C,
(\gh^\C)^*)$-valued superfields.   After these manipulations the
path integral becomes
\begin{equation}
Z = \int [d\tG][d\tH]\,(\det D_{\tA})\,(\det \bar D) e^{-I_\Omega[\tG\tH] +
I_\Omega[\tH]}~,\label{eq:pi2}
\end{equation}
where in the above expression for $D_{\tA}$ it is understood that
$\tA = -D\tH\, \tH^{-1}$.

We will now compute the above ``determinants.''  We will do something
a little bit more general and compute
\begin{equation}
\det D_{\tA}\,\det\bar D_{\bar\tA} \equiv \int
{}[d\tB][d\tC][d\bar\tB][d\bar\tC] e^{-\int_{\Sigma_s} \pair<\bar\tB,
D_{\tA} \bar\tC> - \int_{\Sigma_s} \pair<\tB, \bar
D_{\bar\tA}\tC>}~.\label{eq:dets}
\end{equation}
The above path-integral is determined by the effective action
$W[\tA,\bar\tA]$ defined by
\begin{equation}
\det D_{\tA}\,\det\bar D_{\bar \tA} = e^{-W[\tA,\bar \tA]} \,\det
D\,\det\bar D~.\label{eq:effact}
\end{equation}

There are many ways to compute the effective action.  We choose to
compute it using point-splitting regularisation, or equivalently,
operator product expansions.  The gauge fields $\tA$ and $\bar\tA$
appear linearly in the action in the RHS of (\ref{eq:dets}) coupled to
the currents $\bar\tJ^{\mathrm{gh}}$ and $\tJ^{\mathrm{gh}}$
respectively.  The integrated anomaly $W[\tA,\bar\tA]$ has a nonlocal
series expansion in terms of $\tA$ and $\bar\tA$ obtained by expanding
the terms in the action which contain $\tA$ and $\bar\tA$.  This is a
nonlocal series expansion whose coefficients are correlation functions
of currents.  These correlation functions can be computed using the
operator product expansion, provided that we understand the
currents---which are composite operators---as normal ordered products
regularised using point-splitting, as is usual in two-dimensional
conformal field theory.  This regularisation has the property that the
vacuum expectation value (that is, the one-point function) of any
current $\tJ^{\mathrm{gh}}$ or $\bar\tJ^{\mathrm{gh}}$ vanishes.
Therefore the only contribution to the current-current correlators
comes from the superconformal family of the identity.  However it is
easy to see that the superconformal family of the identity does not
appear (i.e., appears with coefficient zero) in the operator product
expansion of two currents.  We will see this again below, but for now
we simply state that the fermionic and bosonic fields in $\tB$ and
$\tC$ contribute equally but with with opposite signs to the second
order pole, which thus cancels.  In summary, $W[\tA,\bar\tA]$ does not
depend on $(\tA,\bar\tA)$, and in fact,
\begin{displaymath}
\det D_{\tA}\,\det\bar D_{\bar\tA} = \det D\,\det\bar D~.
\end{displaymath}

Inserting this result into (\ref{eq:pi2}) and changing variables
$\tG\mapsto \tG\tH^{-1}$, which has trivial jacobian due to the
absence of gauge anomalies, we arrive at the following expression:
\begin{equation}
Z = \int [d\tG][d\tH][d\tB][d\tC][d\bar\tB][d\bar\tC]
e^{-I_\Omega[\tG] + I_\Omega[\tH]} e^{-\int\pair<\tB,\bar D\tC> +
\pair<\bar\tB,D\bar\tC>}~.\label{eq:pifinally}
\end{equation}

Let us pause for a moment to contemplate our result.  Comparing with
equation (I.4.19), we notice that the only appreciable difference is
the fact that both SWZW sectors have actions corresponding to the same
metric (up to signs).  Furthermore similar arguments to those in Part
I, show that although the three sectors appear to be independent,
there exist constraints that couple them.  Basically one can gauge the
vector subgroup $H$ once again in all three lagrangians,
introducing {\em external\/} gauge superfields and then notice that
the partition function is actually independent of the gauge fields
introduced, which leads to the constraint that the supercurrent which
couples to this gauge superfield has to vanish.  The total
supercurrent has contributions coming from all three SCFTs.

Let us consider the holomorphic sector.  The total current is given by
\begin{displaymath}
\tJ_i^{\mathrm{tot}}(Z) = \tJ_i(Z) + \widetilde \tJ_i(Z) +
\tJ_i^{\mathrm{gh}}(Z)~,
\end{displaymath}
where $\{\tJ_i(Z)\}$ are the subset of the $\gg$ currents (\ref{eq:curr})
corresponding to the subalgebra $\gh$.  The current corresponding to
the gauged sector is given by
\begin{displaymath}
\widetilde \tJ(Z) = - D \tH\tH^{-1}~,
\end{displaymath}
whereas the current corresponding to the ghost sector is defined by
\begin{displaymath}
\tJ_i^{\mathrm{gh}}(Z) = {f_{ij}}^k(\tB_k \tC^j)(Z)~,
\end{displaymath}
with the standard point-splitting convention for the normal
ordering.  These currents will satisfy three commuting current
algebras with the relevant OPEs given by 
\begin{eqnarray}
\tJ_i (Z)\tJ_j (W)& = &\sope[1/1][\Omega_{ij}] + \sope[1/2][{f_{ij}}^k
                      \tJ_k(W)] + \reg~,\nonumber\\  
\widetilde \tJ_i (Z)\widetilde \tJ_j (W)& = &\sope[1/1][-\Omega_{ij}] +
\sope[1/2][{f_{ij}}^k\widetilde \tJ_k (W)] + \reg~,\label{eq:hope}\\
\tJ^{\mathrm{gh}}_i (z)\tJ^{\mathrm{gh}}_j (w) &=&\sope[1/2][{f_{ij}}^k \tJ^{\mathrm{gh}}_k (W)] + \reg~.\nonumber
\end{eqnarray}

Notice that as advertised above, the second order pole of the ghost
currents vanishes, showing that the effective action (\ref{eq:effact})
is indeed independent of the gauge fields.  Adding the central
extensions of the three components of the total conserved current we
see that they cancel each other, which just reiterates the fact that
we have gauged an anomaly-free subgroup.  This guarantees that the
charge which generates the BRST transformations leaving the
``quantum'' action invariant will square to zero.  The theory
resulting from the gauged SWZW model is then defined as the cohomology
of the BRST operator: the states will be the BRST cohomology on
states, and the fields will be the BRST cohomology on fields.  In the
next section we will show that among the BRST invariant fields one
finds the generator of $N{=}1$ superconformal transformations on all
the other BRST invariant fields---and corresponds, in fact, to the one
for the SCFT based on the coset $G/H$.  This result was first obtained
independently in \cite{Rhedin} and in \cite{FSn12}.  As we will see in
Section 7 (see also \cite{FSn12}), if in addition the coset $N{=}1$
superconformal algebra admits an $N{=}2$ extension, the $N{=}2$
generators will also be BRST invariant.

As explained in Part I, the anti-holomorphic sector can be treated
analogously by choosing the anti-holomorphic gauge $\tA=0$.

\section{Coset SCFTs from gauged SWZW models}

In this section we will analyse the quantum field theory which results
after the diagonal gauging of a SWZW model based on a Lie group with a
bi-invariant metric.  We will prove that the resulting theory is a
SCFT and that it can be identified with an $N{=}1$ coset construction.
Indeed, we start by analysing the supersymmetric coset construction in
the case of self-dual Lie algebras.  For the sake of simplicity we
will only consider the holomorphic sector, the treatment of the
antiholomorphic sector being completely analogous.

\subsection{The $N{=}1$ coset construction}

We start with the $N{=}1$ affine algebra $\widehat\gg_{N{=}1}$
(\ref{eq:sope}) based on the self-dual Lie algebra $(\gg,\Omega)$.  We
consider a subalgebra $\gh \subset \gg$, and we fix a basis for it
$\{X_i\}$ which we can think of as a sub-basis of the chosen basis
$\{X_a\}$ for $\gg$.  Then the $N{=}1$ affine algebra $\widehat\gh$ of
the $\{\tJ_i\}$ currents is given by
\begin{displaymath}
\tJ_i(Z) \tJ_j(W) = \sope[1/1][\Omega_{ij}] + \sope[1/2][{f_{ij}}^k
                    \tJ_k(W)] + \reg~,
\end{displaymath}
where $\Omega_{ij}$ are the entries of the restriction $\Omega|_{\gh}$
of $\Omega$ to $\gh\subset\gg$.

Clearly, a coset construction only exists if $\gh$ itself admits a
supersymmetric Sugawara construction based on the above currents.
This means, in light of Section 2, that the bilinear form
$\Omega|_{\gh}$ has to be nondegenerate.  In this case we can
decompose $\gg = \gh \oplus \gh^{\perp}$ which, because of the
invariance of the metric, is not just a decomposition of vector spaces
but also one of $\gh$-modules.  In other words, if we now let
$\{X_{\alpha}\}$ denote a basis for $\gh^{\perp}$, we can summarise
this discussion by saying that $\Omega_{i\alpha}=0$ and that
${f_{i\alpha}}^j=0$.  This fact will play an important role repeatedly
(albeit tacitly at times) in the rest of this paper.

With these remarks behind us, we see that the supersymmetric
energy-mo\-men\-tum tensor corresponding to $\gh$ reads
\begin{displaymath}
\tT_{\gh}(Z) = \half\Omega^{ij}(D\tJ_i\tJ_j)(Z) + \fr{1}{6}
f^{ijk}(\tJ_i(\tJ_j \tJ_k))(Z)~,
\end{displaymath}
and generates an $N{=}1$ superconformal algebra with central charge
\begin{displaymath}
c_{\gh} = {\fr{3}{2}}\dim\gh - {\fr{1}{2}}\Omega_{\gh}^{ij}
          \kappa^{\gh}_{ij}~. 
\end{displaymath}
A straightforward calculation then shows that the coset
energy-momentum tensor defined by
\begin{equation}
\tT_{\gg/\gh} \equiv \tT_{\gg} - \tT_{\gh}\label{eq:tcoset}
\end{equation}
commutes with $\tT_{\gh}$ and generates an $N{=}1$ superconformal
theory with central charge
\begin{equation}
c_{\gg/\gh} \equiv c_{\gg} - c_{\gh} = {\fr{3}{2}}\left(\dim\gg-
              \dim\gh\right) - {\fr{1}{2}} \left(\kappa^{\gg}_{ab}
              \Omega^{ab} - \kappa^{\gh}_{ij}\Omega^{ij}
              \right)~.\label{eq:cosetcc} 
\end{equation}

As explained in Part I, it follows from the structure theorem of
self-dual Lie algebras (see \cite{FSsd} and references therein) that,
just as in the reductive case, there exist natural cosets associated
to nonreductive self-dual Lie algebras.  We hope to return to the
explicit construction of these cosets in a future publication.

\subsection{The SCFTs in the gauged SWZW model}

We will now show that the gauged SWZW model described in the previous
section is a superconformal field theory whose energy momentum tensor
agrees with the coset energy-momentum tensor.  We saw that the quantum
field theory of the gauged WZW model is given by three quantum field
theories coupled by a constraint which we can analyse in the BRST
formalism.  As we now show, each of the three sectors of the theory is
superconformal.  We will also see that the total energy-momentum
tensor is BRST invariant and indeed BRST-equivalent to the $N{=}1$
coset energy momentum tensor in (\ref{eq:tcoset}).

We start then with the SWZW SCFT with group $G$ and metric
$\Omega_{ab}$.  This component corresponds to the original (ungauged)
SWZW model which we discussed in the first part of Section 2.  There
we have seen that we have a set of supercurrents
$\{\tJ_a(Z)\}_{a=1}^{\dim\gg}$ whose OPE is given by (\ref{eq:sope}).
We have also seen that according to the $N{=}1$ self-dual Sugawara
construction this SWZW sector does give rise to a SCFT if and only if
this metric is nondegenerate.  In this case, the energy-momentum
tensor
\begin{equation}
\tT_{\gg}(Z) = \half\Omega^{ab}(D\tJ_a\tJ_b)(Z) + \fr{1}{6} f^{abc}
(\tJ_a (\tJ_b \tJ_c))(Z)~,\label{eq:tgg}
\end{equation}
obeys a Virasoro algebra with the central charge
\begin{equation}
c_{\gg} = {\fr{3}{2}}\dim\gg - {\fr{1}{2}}\Omega^{ab}\kappa^{\gg}
_{ab}~.\label{eq:cgg}
\end{equation}

The next ingredient is provided by the SWZW model with group $H\subset
G$ and metric $-\Omega_{ij}$.  This is characterised by the set of
currents $\{\widetilde\tJ_i(Z)\}_{i = 1}^{\dim\gh}$ whose OPE is given
by (\ref{eq:hope}).  Applying (once again) the argument of Section 2, we
get that this current algebra gives rise to a SCFT if and only if the
restriction of $\Omega$ to $\gh$ is itself nondegenerate.  In this
case the corresponding energy-momentum tensor
\begin{equation}
\widetilde \tT_{\gh}(Z) =
-\half\Omega^{ij}(D\widetilde\tJ_i\widetilde\tJ_j)(Z) +
 \fr{1}{6} f^{ijk}(\widetilde\tJ_i(\widetilde\tJ_j \widetilde\tJ_k))(Z)~,
\label{eq:tgh}
\end{equation}
will generate an $N{=}1$ superconformal algebra with central charge
\begin{equation}
\widetilde c_{\gh} = {\fr{3}{2}}\dim\gh + {\fr{1}{2}}\Omega^{ij}
                 \kappa^{\gh}_{ij}~.\label{eq:cgh}  
\end{equation}

The last sector of the theory consists of a set of $(\dim\gh)$
supersymmetric $(\tB,\tC)$ systems of superconformal weights
$(\fr{1}{2},0)$ respectively, with OPE given by
\begin{displaymath}
\tB_i(Z)\tC^j(W) = \sope[1/2][{\delta_i}^j] + \reg~.
\end{displaymath}
The energy-momentum tensor for this $(\tB,\tC)$ system has the
standard form
\begin{equation}
\tT_{\mathrm{gh}}(Z) = -\half\left(\tB_i\d\tC^i\right)(Z) +
                   \half\left(D\tB_i D\tC^i\right)(Z)~,\label{eq:tghost}
\end{equation}
and obeys the superconformal algebra with central charge
\begin{equation}
c_{\mathrm{gh}} = -3\dim\gh~.\label{eq:cghost}
\end{equation}

Finally, we introduce the last ingredient of this theory: the BRST
current:
\begin{equation}
j(Z) = (\tC^i( \tJ_i+\widetilde \tJ_i+\half
                         \tJ^{\mathrm{gh}}_i))(Z)~.\label{eq:jbrst}
\end{equation}
It follows from the formulae in Appendix A, that the BRST variation
$d\tA(Z)$ of a superfield $\tA(Z)$ is given by:
\begin{displaymath}
d\tA \equiv [\![j,\tA]\!]_{\shalf}~.
\end{displaymath}
It follows using the associativity axiom that
\begin{displaymath}
d^2\tA = \half [\![[\![j,j]\!]_{\shalf},\tA]\!]_{\shalf}~;
\end{displaymath}
whence, in the absence of any further relations, $d^2=0$ provided that
$[\![j,j]\!]_{\shalf} = D\tX$ for some $\tX$.  Indeed it is an
easy computation using the formulae in Appendix A, that actually
$[\![j,j]\!]_{\shalf} = 0$, which simply reiterates the fact that
we have gauged an anomaly-free subgroup and justifies our assuming the
absence of gauge anomalies.

\subsection{The energy-momentum tensor}

The total energy-momentum tensor is given by the sum of the three
commuting terms given by (\ref{eq:tgg}), (\ref{eq:tgh}), and (\ref{eq:tghost}):
\begin{displaymath}
\tT(z) = \tT_{\gg}(z) + \widetilde \tT_{\gh}(z) +
\tT_{\mathrm{gh}}(z)~,
\end{displaymath}
whose central charge is obtained by adding up (\ref{eq:cgg}),
(\ref{eq:cgh}), and (\ref{eq:cghost}):
\begin{displaymath}
c = \left({\fr{3}{2}}\dim\gg - \half\Omega^{ab}\kappa^{\gg}_{ab}
    \right) - \left({\fr{3}{2}}\dim\gh - \half\Omega^{ij}
    \kappa^{\gh}_{ij}\right)~. 
\end{displaymath}
Notice that this agrees with the coset central charge given by
equation (\ref{eq:cosetcc}).  This prompts us to compare $\tT(z)$ with
the energy-momentum tensor of the corresponding coset construction.
We introduce for this purpose
\begin{displaymath}
\tT_{\gh} = \half\Omega^{ij}(D\tJ_i\tJ_j) + \fr{1}{6} f^{ijk}(\tJ_i(\tJ_j
\tJ_k))~.
\end{displaymath}
Our total energy-momentum tensor then splits into a sum of two
{\em commuting\/} terms
\begin{displaymath}
\tT(z) = \tT_{\gg/\gh}(z) + \tT'(z)~,
\end{displaymath}
with $\tT_{\gg/\gh}(z)$ being the coset energy-momentum tensor defined
by (\ref{eq:tcoset}) and
\begin{displaymath}
\tT'(z) = \tT_{\gh}(z) + \widetilde \tT_{\gh}(z) + \tT_{\mathrm{gh}}(z)~.
\end{displaymath}
Moreover, a short computation shows us that $\tT'$ satisfies an
$N{=}1$ superconformal algebra with vanishing central charge, $c'=0$. 

Our aim now is to show that $\tT$, $\tT_{\gg/\gh}$ and $\tT'$ are
BRST-invariant, so that they are physical operators (that is, they
induce operators in the physical space).  Moreover we will show that
$\tT'$ is BRST-trivial, which means that it acts trivially on physical
states.

To this effect it is convenient to list the following identities:
\begin{eqnarray*}
d\tB_i(Z) &=& \tJ^{\mathrm{tot}}_i(Z)~,\\
d\tC^i(Z) &=& \half{f_{jk}}^i\tC^j\tC^k(Z)~,\\
d\tJ_a(Z) &=& \Omega_{aj}D\tC^j(Z) + {f_{aj}}^b\tC^j\tJ_b(Z)~,\\
d\widetilde\tJ_i(Z) &=& -\Omega_{ij}D\tC^j(Z) + {f_{ij}}^k
\tC^j\widetilde \tJ_k(Z)~.
\end{eqnarray*}
Using these relations we deduce the following
\begin{displaymath}
d\tJ^{\mathrm{gh}}_i(Z) = {f_{ij}}^k(\tJ_k + \widetilde\tJ_k)\tC^j -
      {f_{ik}}^m {f_{j\ell}}^k \tB_m (\tC^j\tC^\ell)~.
\end{displaymath}
We can now use these identities and the fact that $d$ is an odd
derivation over the normal ordered product, to prove that:
\begin{displaymath}
d\tT(Z) = d\tT_{\gg/\gh}(Z) = d\tT'(Z) = 0~;
\end{displaymath}
that is, $\tT(Z)$, $\tT_{\gg/\gh}(Z)$ and $\tT'(Z)$ are
BRST-invariant.  Furthermore, it also follows that there exists an
operator
\begin{eqnarray*}
\tO(Z) &=& -\fr{1}{4} \Omega^{ij} D\tB_i(\tJ_j - \widetilde \tJ_j) + 
 \fr{1}{4} \Omega^{ij} \tB_i(D\tJ_j - D\widetilde \tJ_j)\\
& & \mbox{} +  \fr{1}{6} f^{ijk} \tB_i (\tJ_j\tJ_k +
\widetilde\tJ_j\widetilde\tJ_k - \tJ_j\widetilde\tJ_k)
\end{eqnarray*}
such that
\begin{displaymath}
\tT'(Z) = d\tO(Z)~;
\end{displaymath}
in other words, $\tT'(Z)$ is BRST trivial, whence it is zero in
cohomology.

These results imply that the quantum field theory defined by the
gauged SWZW model (\ref{eq:pifinally}) is superconformal, with
energy-momentum tensor given by $\tT(Z)$ and its antiholomorphic
counterpart $\bar \tT(\bar Z)$.  Moreover $\tT(Z)$ (and similarly for
$\bar \tT(\bar Z)$) is precisely the coset energy-momentum tensor
$\tT_{\gg/\gh}(Z)$; whence we conclude that the gauged SWZW
model defines a superconformal field theory which can be identified
with the supersymmetric coset $G/H$.

\section{The SWZW model in components}

In the last three sections we have seen that it is possible to
consistently define the $N{=}1$ supersymmetric WZW model associated
to a general self-dual Lie group (both gauged and ``as is'') in
superspace, that it yields a superconformal invariant theory, and
moreover we have exactly determined the SCFT it describes.  While this
is certainly a satisfactory state of affairs, we are motivated to
consider the formulation of the SWZW model in components because of
two main reasons.   On the one hand, there has been quite a lot of
work done in components and we think it is useful to clarify the
correspondence between the two approaches; especially when it comes to
the gauged SWZW model.  And on the other hand, in the discussion on
$N{=}2$ Kazama--Suzuki cosets in Section 7, we will be forced to work
in components, since there is no local expression of the extra
supersymmetry generators in terms of superfields.

We start this section by exhibiting the underlying Lie supergroup
structure in the SWZW model.  To be precise the superfield
$\tG(Z,\bar Z)$ does not take values in the Lie group $G$ but rather
in a Lie supergroup with body $G$.  This Lie supergroup possesses an
invariant metric and the symmetries of the SWZW model arise as
symmetries of this structure in a manifest manner.  Moreover the
underlying supergroup will also suggest a natural parametrisation of
the superfield $\tG(Z,\bar Z)$ into components.

\subsection{The underlying Lie supergroup}

The superfield $\tG(Z,\bar Z)$ can be parametrised in several ways.  A
parametrisation which teaches us where $\tG$ really takes values, is the
following.  Let $\tX(Z,\bar Z)$ be a superfield given by
\begin{displaymath}
\tX(Z,\bar Z) = x(z,\bar z) + \theta \chi(z,\bar z) + \bar\theta
\bar\chi(z,\bar z) + \theta\bar\theta f(z,\bar z)~,
\end{displaymath}
where $x,f$ are bosonic and $\chi,\bar\chi$ are fermionic fields with
values in the Lie algebra $\gg$ of $G$.  It would be tempting 
to call $\tX$ a $\gg$-valued superfield, but because of the $\theta$'s
this is strictly speaking not correct.  Instead, $\tX$ takes values in a
Lie superalgebra $\gs$ built out of $\gg$ in the following canonical
fashion.  As a vector space $\gs$ consists of four copies of $\gg$.
If we write $\gs_{\bar 0}$ and $\gs_{\bar 1}$ for the even and odd
subspaces of $\gs$, then as vector spaces both are isomorphic to
$\gg\oplus\gg$. If $\tX$ is as above then $(x,f) \in \gs_{\bar 0}$ and
$(\chi,\bar\chi)\in \gs_{\bar 1}$.  The Lie bracket of $\gs$ is
induced from the one in $\gg$.  Indeed, if $\tX$ is as above and $\tY
= y + \theta\gamma + \bar\theta\bar\gamma + \theta\bar\theta g$, then
\begin{eqnarray*}
\mbox{}[\tX,\tY] &=& [x,y] + \theta \left( [x,\gamma] -
[y,\chi]\right) + \bar\theta \left( [x,\bar\gamma] -
[y,\bar\chi]\right)\\
& &\mbox{} + \theta\bar\theta \left( [x,g] - [y,f] - [\chi,\bar\gamma]
+ [\bar\chi, \gamma] \right)~,
\end{eqnarray*}
whence we see that, for instance, $\gg \subset\gs_{\bar 0}$ is
naturally a subalgebra, and that in fact $\gs_{\bar 0}$ is the abelian
extension $\gg \ltimes \gg_{\mathrm{ab}}$, where $\gg$ acts on
$\gg_{\mathrm{ab}}$ via the adjoint representation, and
$\gg_{\mathrm{ab}}$ is $\gg$ made abelian.  Similarly $\gg\subset
\gs_{\bar 0}$ acts on $\gs_{\bar 1}$ via two copies of the adjoint
representation.

The metric on $\gg$ induces a metric on $\gs$ in the following way.
if $\tX$ and $\tY$ are as above, we write
\begin{eqnarray*}
\lefteqn{\pair<\tX,\tY> = \pair<x,y> + \theta \left( \pair<x,\gamma> -
\pair<y,\chi>\right)  + \bar\theta \left( \pair<x,\bar\gamma> -
\pair<y,\bar\chi>\right)}\\
& & {} + \theta\bar\theta \left( \pair<x,g> - \pair<y,f> -
\pair<\chi,\bar\gamma> + \pair<\bar\chi, \gamma> \right)~.
\end{eqnarray*}
Each component is $\gs$-invariant, but only the
$\theta\bar\theta$-component is nondegenerate.  Notice that this
component is precisely what is projected out by the Berezin integral
$\int_B \pair<\tX,\tY>$.  In other words the composition of Berezin
integral and the metric on $\gg$ define a metric on $\gs$ which is
clearly invariant.  In other words, $\gs$ is self-dual.

As discussed in Part~I (see also \cite{FSsd}) any indecomposable
self-dual Lie algebra which is not simple or one-dimensional is
obtained as a double extension of a self-dual Lie algebra by either a
one-dimensional or a simple Lie algebra.  As remarked in \cite{FSsd},
although double extensions still produce self-dual Lie superalgebras,
there exists no proof of (nor a counterexample against) the similar
result for self-dual Lie superalgebras.  We can check in this case
that $\gs$ is a double of extension of the abelian Lie superalgebra
$\gs_{\bar 1} = \gg \oplus \gg$ with off-diagonal metric:
\begin{displaymath}
\pair<(\chi,\bar\chi),(\gamma,\bar\gamma)> =
\pair<\bar\chi,\gamma> - \pair<\chi,\bar\gamma>~,
\end{displaymath}
by the Lie algebra $\gg$, acting under two copies of the adjoint
representation.

If we now parametrise $\tG(Z,\bar Z) = \exp \tX(Z,\bar Z)$ we see that
the superfield $\tG(Z,\bar Z)$ describing the SWZW model in equation
(\ref{eq:SWZW}), takes values in a Lie supergroup $S$ with Lie
superalgebra $\gs$.  It now becomes obvious that the SWZW action
(\ref{eq:SWZW}) possesses the symmetries ascribed to it in Section 2.
Just as the symmetries of the ordinary WZW model follow from the fact
that $G$ is a self-dual Lie group, the symmetries of the SWZW model
follow from the fact that $S$ is a self-dual Lie supergroup.

Since the Lie supergroup $S$ has $G$ as its body, it follows from the
structure theory of Lie supergroups that we can decompose elements in
$S$ as follows:
\begin{displaymath}
S = G \cdot S_N
\end{displaymath}
where $S_N$ are those group elements in $S$ of the form $\1 +
\hbox{nilpotent}$.  In other words, elements of $S_N$ can be
understood as exponentials of elements $\tX_N$ of $\gs$ of the form
$\tX_N = \theta\lambda + \bar\theta\bar\lambda + \theta\bar\theta f$.
Hence this decomposition suggests a different parametrisation of the
superfield $\tG(Z,\bar Z)$:
\begin{equation}
\tG = g \, \tG_N = g \left[ 1 + \theta\lambda + \bar\theta\lambda +
\theta\bar\theta (f -\half \lambda\bar\lambda + \half
\bar\lambda\lambda)\right ]~,\label{eq:precomp}
\end{equation}
where $g$ takes values in the Lie group $G$, and $\lambda$,
$\bar\lambda$ and $f$ all take values in the Lie algebra.  Notice
however that because $\lambda$ and $\bar\lambda$ are odd, what appears
in the $\theta\bar\theta$-term is not the Lie bracket
$[\lambda,\bar\lambda]$ but the anticommutator.  Hence the
$\theta\bar\theta$-component of $\tG_N$ does not live in the Lie
algebra $\gg$ but in $\gg \oplus S^2\gg$.

\subsection{The component action for the SWZW model}

In order to write the SWZW action in components we start with
(\ref{eq:SWZW}) and parametrise the superfield $\tG$ in a more symmetric
version of (\ref{eq:precomp}):
\begin{eqnarray}
\tG &=& g \left[ 1 + \theta \psi + \bar\theta g^{-1}\bar\psi g +
     \theta\bar\theta (a-\psi g^{-1}\bar\psi g) \right]~,\nonumber\\
\tG^{-1} &=& \left[ 1 - \theta \psi - \bar\theta g^{-1}\bar\psi g -
          \theta\bar\theta (a-g^{-1}\bar\psi g\psi) \right] g^{-1}~, 
                                                        \label{eq:par}
\end{eqnarray}
where $g:\Sigma\to G$ and all the other fields are defined on $\Sigma$
and take values in the Lie algebra $\gg$.  As usual in supersymmetric
$\sigma$-models, the fermions are sections of the spinor bundle on
$\Sigma$ twisted by the pull-back of the tangent bundle of the Lie
group $TG$.  But here, in addition, there is an implicit
trivialisation of the tangent bundle of the Lie group, so that $\psi$
and $\bar\psi$ are actually Lie algebra valued.  This trivialisation
is accomplished via left or right translations.  One choice is natural
for $\psi$ and another for $\bar\psi$ which explains the seemingly
asymmetrical way in which they appear in (\ref{eq:par}).  We will have
more to say about this parametrisation below.

Introducing the expression of the superfield (\ref{eq:par}) in the SWZW
action, eliminating the auxiliary field $a$ by its equation of motion
$a=0$, and after some algebra we obtain:
%\begin{eqnarray}
%I_\Omega[g,\psi,\bar\psi,a] &= 
%  &\int_{\Sigma}\pair<g^{-1}\d g,g^{-1}\bar\d g> +
%   \int_{B}\pair<\tilde g^{-1}\d_t \tilde g, \left[\tilde g^{-1}\d
%\tilde g, \tilde g^{-1}\bar\d \tilde g\right]> \nonumber\\
%  &&- \int_{\Sigma} \pair<\psi , \bar\d\psi + [g^{-1}\bar\d
%   g,\psi]> + \pair<\bar\psi , \d\bar\psi - [\d
%   g~g^{-1},\bar\psi]>\nonumber\\ 
%  &&+ \int_{\Sigma} \pair<a,a>~.\label{eq:comp1}
%\end{eqnarray}                       
\begin{equation}
I_\Omega[g,\psi,\bar\psi] = 
 I_\Omega[g] - \int_{\Sigma} \pair<\psi ,
\bar\nabla\psi> + \pair<\bar\psi , \nabla\bar\psi>~,\label{eq:compwzw} 
\end{equation}
where $I_\Omega[g]$ is the WZW action given by (I.2.11), and the last
two terms describe describe Weyl fermions in the adjoint
representation of $\gg$, axially coupled to the bosonic field $g$ via
the covariant derivatives:
\begin{eqnarray*}
\nabla &\equiv& \ad_g \circ \d \circ \ad_g^{-1} = \d - [\d g~g^{-1},-]~,\\
\bar\nabla &\equiv& \ad_g^{-1} \circ \bar\d \circ \ad_g =  \bar\d +
{}[g^{-1}\bar\d g,-]~,
\end{eqnarray*}
with $\ad_g$ denoting the adjoint action of the group $G$ on its Lie
algebra $\gg$:
\begin{displaymath}
\ad_g (X) = g\,X\,g^{-1}~.
\end{displaymath}

The above action is clearly supersymmetric.  Indeed, one can easily
check that (\ref{eq:compwzw}) is invariant under the following
supersymmetry transformations:
\begin{eqnarray*}
\delta g &=& \epsilon g\psi + \bar\epsilon \bar\psi g~,\\
\delta \psi &=& \epsilon (g^{-1}\d g - \psi^2) - \bar\epsilon
               [\psi,g^{-1}\bar\psi g]~,\\
\delta \bar\psi &=& \epsilon [g\psi g^{-1},\bar\psi] + \bar\epsilon
                   (\bar\d gg^{-1} + {\bar\psi}^2)~.
\end{eqnarray*}

In addition, the SWZW action (\ref{eq:compwzw}) is invariant under both
bosonic and fermionic symmetry transformations which can be determined
either by breaking (\ref{eq:swzwinv}) into components, or by direct
investigation of (\ref{eq:compwzw}).  The bosonic symmetry
transformations read
\begin{eqnarray}
g(z,\bar z) &\mapsto& h^{-1}(z)g(z,\bar z){\bar h}(\bar
z)\label{eq:bos1}\\
\psi(z,\bar z) &\mapsto& {\bar h}^{-1}(\bar z)\psi(z,\bar z){\bar
h}(\bar z)\label{eq:bos2}\\
\bar\psi(z,\bar z) &\mapsto& h^{-1}(z)\bar\psi(z,\bar
z)h(z)~,\label{eq:bos3}
\end{eqnarray}
with $h$ and $\bar h$ being holomorphic and antiholomorphic maps,
respectively, from $\Sigma$ to the group $G$.  The bosonic conserved
currents associated to this invariance are
\begin{displaymath}
I(z) = -(\d g~g^{-1} + g\psi^2 g^{-1})~, \qquad
\bar I(z) = (g^{-1}\bar\d g - g^{-1}{\bar\psi}^2 g)~,
\end{displaymath}
with the equations of motion $\bar\d I=\d\bar I=0$.  Notice the
structure of these currents: the first term is nothing but the
conserved current of the bosonic WZW model, and they have a second
term due to the fermions, which depends on the structure constant of
the algebra $\gg$.  The fermionic symmetries leave $g$ inert and act
on the fermions as follows:
\begin{eqnarray}
\psi(z,\bar z) &\mapsto& \psi(z,\bar z) + g^{-1}(z,\bar z)\chi(z)
                g(z,\bar z)\label{eq:fer1}\\  
\bar\psi(z,\bar z) &\mapsto& \bar\psi(z,\bar z) + g(z,\bar z)
               \bar\chi(\bar z)g^{-1}(z,\bar z)~,\label{eq:fer2}
\end{eqnarray}
with $\chi$, $\bar\chi$ holomorphic and antiholomorphic maps,
respectively, from $\Sigma$ to $\gg$.  The fermionic conserved
currents corresponding to this invariance are
\begin{displaymath}
\Psi (z) = -g\psi g^{-1}~, \qquad
\bar\Psi (\bar z) = g^{-1}\bar\psi g~,
\end{displaymath}
with the equations of motion $\bar\d \Psi=\d\bar\Psi=0$.

One can easily verify that the above conserved currents coincide with
the components of the supercurrents $\tJ$, $\bar\tJ$ (after imposing
the equations of motion), that is
\begin{displaymath}
\tJ (Z) = \Psi(z) + \theta I(z)~, \qquad
\bar\tJ (\bar Z) = \bar\Psi(\bar z) + \bar\theta \bar I(\bar z)~.
\end{displaymath}
Therefore it should not come as a surprise that the current algebra
generated by these currents coincides with the one obtained by
breaking (\ref{eq:sope}) into components.  Indeed, if we take $I$, $\bar I$
and $\Psi$, $\bar\Psi$ as the dynamical variables and compute the
fundamental Poisson brackets we find:
\begin{eqnarray*}
\left\{ I_a (z),I_b (w)\right\} &=& \left( \Omega_{ab}\d_w 
+ {f_{ab}}^c I_c(w)\right)\delta (z-w)\\
\left\{ I_a (z),\Psi_b (w)\right\} &=& {f_{ab}}^c \Psi_c(w) \delta
                                      (z-w)\\ 
\left\{ \Psi_a (z),\Psi_b (w)\right\} &=& \Omega_{ab}\delta
(z-w)~,
\end{eqnarray*}
which upon quantisation become:
\begin{eqnarray}
I_a (z)I_b (w) &=& \ope[2][\Omega_{ab}] + \ope[1][{f_{ab}}^c I_c (w)]
+ \reg\label{eq:ktcb1}\\
I_a (z)\Psi_b (w) &=& \ope[1][{f_{ab}}^c \Psi_c (w)] +
\reg\label{eq:ktcb2}\\
\Psi_a (z)\Psi_b (w) &=& \ope[1][\Omega_{ab}] + \reg~,\label{eq:ktcb3}
\end{eqnarray}
and similar formulas for the antiholomorphic currents.

\subsection{The quantum theory}

The quantum theory is described by the path integral
\begin{equation}
Z = \int [dg][d\psi][d\bar\psi] e^{-I_{\Omega}[g] + \int 
    \pair<\psi , \bar\nabla\psi> + \pair<\bar\psi ,
    \nabla\bar\psi>}~.\label{eq:path} 
\end{equation}

One can of course decouple the fermions in (\ref{eq:path}) by performing an
axial gauge transformation; this will incur in a nontrivial jacobian
in the path integral which is described by the effective action, $W[g]
= - I_{\shalf\kappa}[g]$, which is nothing but a WZW action on $G$,
with a metric proportional to the Killing metric on $\gg$.  Thus, in
terms of the free fermions the path integral reads
\begin{equation}
Z = \int [dg][d\psi][d\bar\psi] e^{-I_{\Omega-\shalf\kappa}[g] +
    \int \pair<\psi , \bar\d\psi> + \pair<\bar\psi ,
    \d\bar\psi>}~.\label{eq:decpath}
\end{equation}

As shown in Part I (see also \cite{FSsd}) the bosonic action is still
generically unconstrained since for a generic self-dual Lie algebra
$\Omega - \half \kappa$ is nondegenerate (see Theorem I.3.6 for the
precise statement).  The proof of this result given in Part I relies
strongly on the structure theory of self-dual Lie algebras.  In the
light of the above discussion and of equation (\ref{eq:decpath}), a
heuristic argument for the nondegeneracy of $\Omega-\half\kappa$ might
be obtained simply by supersymmetrising the WZW model, and then
decoupling the fermions.

At the level of the current algebra, decoupling the fermions amounts
to a redefinition of the bosonic current.  In the last paragraph we
have written the $N{=}1$ current algebra in terms of a basis of
currents $\{I_a,\Psi_a\}$.  If we now redefine the bosonic current
\begin{equation}
J_a \equiv I_a - \half\Omega^{bd}{f_{ab}}^c (\Psi_c\Psi_d)~,\label{eq:deccurr}
\end{equation} 
and leave the fermionic current unmodified, then the fermions
decouple:
\begin{eqnarray}
J_a (z)J_b (w) &=& \ope[2][\Omega_{ab}-\half\kappa_{ab}] +
                  \ope[1][{f_{ab}}^c J_c (w)] + \reg~,\label{eq:bca} \\
J_a (z)\Psi_b (w) &=& \reg~,\label{eq:bca2} \\
\Psi_a (z)\Psi_b (w) &=& \ope[1][\Omega_{ab}] + \reg~.\label{eq:bca3}
\end{eqnarray}
In other words the modified bosonic currents commute with the
fermionic ones, while the residue of the double pole in the OPE of the
bosonic current with itself receives a shift proportional to the
Killing metric.  Notice that the central extension in the OPE of the
bosonic currents corresponds exactly to the metric of the bosonic part
of the SWZW action, and similarly the central extension of the
fermionic OPE corresponds to the metric of the fermionic part of the
action.

This decoupled basis is particularly convenient for writing the
superconformal algebra in components.  Indeed if we define the
components of the supersymmetric generator of the superconformal
algebra by
\begin{displaymath}
\tT(Z) = \half \sG(z) + \theta \sT(z)~,
\end{displaymath}
then we get
\begin{eqnarray*}
\sT(z) &=& \half\Omega^{ab}(J_a J_b)(z) +
        \half\Omega^{ab}(\d\Psi_a\Psi_b)(z)~,\\
\sG(z) &=& \Omega^{ab}(J_a \Psi_b)(z) - \fr{1}{6} f^{abc}
        (\Psi_a(\Psi_b\Psi_c))(z)~.
\end{eqnarray*}
Notice that in this basis the energy-momentum tensor $\sT(z)$ is
written as a sum of two independent terms, the first one  being the
bosonic Sugawara energy-momentum tensor corresponding to the bosonic
current algebra (\ref{eq:bca}), whereas the second one is the standard
energy-momentum of $\dim G$ free fermions.

Let us conclude this section with a brief discussion on the choice of
parametrisation made at the beginning of this section.  It is clear
that the parametrisation chosen for the superfield $\tG$ in
(\ref{eq:par}) is not unique.  Indeed, one can redefine the fermionic
fields in such a way that the fermions in the SWZW action end up
either coupled or uncoupled to the bosonic field $g$.  This naturally
raises the question whether these different parametrisations are
equivalent or not, and if they are not, how does one choose the
``right'' parametrisation.

At the classical level, there is no real distinction between them.
Indeed, let us consider for concreteness the following two
parametrisations: the one defined by (\ref{eq:par}) which gives rise to the
coupled fermions in the (component) SWZW action, and the
parametrisation obtained from (\ref{eq:par}) by performing the following axial
gauge transformation
\begin{displaymath}
\psi \mapsto g^{-1}\psi g\qquad\hbox{and}\qquad \bar\psi \mapsto g\psi
g^{-1}~.
\end{displaymath}
This yields a component action which is simply the sum of the bosonic
action and the free fermions, and which is thus superconformally
invariant.  

The situation at the quantum level is slightly different, and the free
and the coupled parametrisations are no longer equivalent.  The reason
being that the change of variables which decouple the fermions
classically, gives rise to a nontrivial jacobian at the quantum level.
More explicitly one can see this by simply comparing the ``decoupled''
path integral (\ref{eq:decpath}) with the path integral corresponding to the
``free'' parametrisation:
\begin{equation}
Z' = \int [dg][d\psi][d\bar\psi] e^{-I_{\Omega}[g] +
    \int \pair<\psi , \bar\d\psi> + \pair<\bar\psi ,
    \d\bar\psi>}~.\label{eq:freez}  
\end{equation}
The two parametrisations give rise to two different quantum theories
characterised by different metrics in the bosonic part of the action. 

Which is then the correct parametrisation?  Keeping in mind that we
are interested in finding a lagrangian description for certain classes
of SCFTs, a natural choice of parametrisation is one which yields a
(manifestly) superconformal theory; in other words, a SWZW model whose
symmetry algebra is an $N{=}1$ affine Lie algebra.  We have already
seen that, by using the coupled parametrisation (\ref{eq:par}), we
obtain the right current algebra.  On the other hand, if we would take
the path integral (\ref{eq:freez}) and we write down the corresponding
current algebra we see immediately that it does not agree with the N=1
affine algebra which we need for the Sugawara construction.  (In the
decoupled basis, both the bosonic and the fermionic OPEs have the
unshifted metric as central term.)  In summary, this justifies our
choice of parametrisation of the superfield $\tG$ in (\ref{eq:par}).

\section{The gauged SWZW models in components and Witten's action}

We have obtained two different actions for the SWZW model: in
superfields and in components.  Moreover we have obtained a superfield
action for the gauged SWZW model.  We can obtain a component action in
either of two ways: we can gauge the component action (\ref{eq:compwzw})
or we can break the superfield gauged action (\ref{eq:gaugact}) into
components.  In this section we show that both methods yield
equivalent actions, and that these actions are in turn equivalent to
the action written down by Witten in \cite{Witten}, in his formulation
of the topological Kazama--Suzuki models.

\subsection{Gauging the component action}

We now set out to gauge {\em both\/} the fermionic
(\ref{eq:fer1})-(\ref{eq:fer2}) and bosonic (\ref{eq:bos1})-(\ref{eq:bos3})
symmetries of the component action of the SWZW model.  In other words,
we will now construct an extension of the action (\ref{eq:compwzw})
which is invariant under bosonic transformations of the form
\begin{eqnarray*}
g(z,\bar z) &\mapsto& \lambda^{-1}(z,\bar z)g(z,\bar z) \lambda(z,\bar
                     z)\\
\psi(z,\bar z) &\mapsto& \lambda^{-1}(z,\bar z)\psi(z,\bar z)
                        \lambda(z,\bar z)\\
\bar\psi(z,\bar z) &\mapsto& \lambda^{-1}(z,\bar z)\bar\psi(z,\bar z)
                            \lambda(z,\bar z)~,
\end{eqnarray*}
and also under the fermionic transformations
\begin{eqnarray*}
\psi(z,\bar z) &\mapsto& \psi(z,\bar z) + g^{-1}(z,\bar z)\chi(z,\bar z)
                        g(z,\bar z)\\  
\bar\psi(z,\bar z) &\mapsto& \bar\psi(z,\bar z) + g(z,\bar z)
                            \bar\chi(z,\bar z)g^{-1}(z,\bar z)~,
\end{eqnarray*}
and leaving $g$ inert.

We will be concerned with infinitesimal gauge transformations.  To
this end, let parametrise $\lambda=e^{\omega}$ and consider $\chi$ and
$\bar\chi$ as infinitesimal parameters.  The infinitesimal gauge
transformations now read:
\begin{eqnarray*}
\delta g &=& [g,\omega]\\
\delta\psi &=& g^{-1}\chi g + [\psi,\omega]\\
\delta\bar\psi &=& g\bar\chi g^{-1} + [\bar\psi,\omega]~.
\end{eqnarray*}

Since the bosonic field $g$ is inert under the fermionic symmetry, the
gauging of the bosonic part of the action is simply the gauged WZW
model given by (I.4.5).  We therefore focus on the fermionic part of
the action (\ref{eq:compwzw}).

Because of factorisation, we will consider each term separately.  We
will apply the Noether method to the action:
\begin{displaymath}
S^{(0)} = \int_{\Sigma} \pair<\psi,\bar\nabla\psi>~,
\end{displaymath}
whose variation under an infinitesimal gauge transformation reads
\begin{displaymath}
\delta S^{(0)} = 2\int \pair<\bar\d\omega,g\psi^2 g^{-1}> -
                 \pair<\bar\d\chi, g\psi g^{-1}>~.
\end{displaymath}
We introduce at this point a bosonic gauge field $\bar A$ and a
fermionic gauge field $\bar\sigma$, a $(0,1)$ form and a
$(\half,1)$-form respectively, whose variations under an infinitesimal
gauge transformation are given by
\begin{eqnarray*}
\delta\bar A &=& \bar\d\omega + [\bar A,\omega]~,\\
\delta\bar\sigma &=& \bar\d\chi + [\bar A,\chi] +
                    [\bar\sigma,\omega]~.
\end{eqnarray*}

Following the Noether procedure we construct the first order
correction to the $S^{(0)}$ to be equal to
\begin{displaymath}
S^{(1)} = - 2 \int \pair<\bar A,g\psi^2 g^{-1}> -
          \pair<\bar\sigma,g\psi g^{-1}>~.
\end{displaymath}
This will cancel the two terms in $\delta S^{(0)}$ but will yield
further 
\begin{equation}
\delta (S^{(0)} + S^{(1)}) = 2 \int
\pair<\bar\sigma,\chi>~.\label{eq:prob}
\end{equation}

Now we face an obstruction, because one can easily see that there are
no further local terms which can be added whose infinitesimal gauge
transformation would cancel (\ref{eq:prob}).  One way around this
problem is to parametrise the fermionic gauge field $\bar\sigma$ in
terms of a new fermionic field $\eta$, a $(\half,0)$-form, such that
\begin{equation}
\bar\sigma = \bar\d_{\bar A} \eta = \bar\d\eta [\bar
A,\eta]~,\label{eq:bsigma}
\end{equation}
where we have introduced the covariant derivative $\bar \d_{\bar A} =
\bar\d + [\bar A,-]$.  The infinitesimal gauge transformation of
$\eta$ is fixed by that of $\sigma$ to be:
\begin{displaymath}
\delta\eta = \chi + [\eta,\omega]~.
\end{displaymath}
Then we can add the following term to the action:
\begin{displaymath}
S^{(2)} = - \int \pair<\bar\sigma,\eta>~.
\end{displaymath}
The resulting action $S = S^{(0)} + S^{(1)} + S^{(2)}$ is indeed gauge
invariant.  In fact, it will be convenient to slightly modify this
action so that (\ref{eq:bsigma}) will appear naturally as an equation of
motion.  In that case the gauge invariant fermionic action will take
the form
\begin{displaymath}
S = \int \pair<g\psi g^{-1},\bar \d_A(g\psi g^{-1})> + 2
    \pair<\bar\sigma,g\psi g^{-1}> - 2\pair<\bar\sigma,\eta> -
    \pair<\eta,\bar \d_A\eta>~. 
\end{displaymath}

One can proceed in a similar fashion and gauge the other fermionic
term in (\ref{eq:compwzw}), introducing the corresponding gauge fields,
$A$, $\sigma$ and $\bar\eta$ (with $\sigma=\d_{A}\bar\eta$), which
transform like
\begin{eqnarray*}
\delta A &=& \d_A\omega ~,\\
\delta\sigma &=& \d_A\bar\chi + [\sigma,\omega]~, \\ 
\delta\bar\eta &=& \d\bar\chi + [\bar\eta,\omega]~.
\end{eqnarray*}
Then the full action of the gauged WZW model will be given by
\begin{eqnarray}
I &= &I_B[g,A,\bar A]\nonumber\\
    &&- \int_{\Sigma} \pair<g\psi g^{-1},\bar \d_{\bar A}(g\psi g^{-1})> +
     \pair<g^{-1}\bar\psi g,\d_A(g^{-1}\bar\psi g)>\nonumber\\
    &&+ \int_{\Sigma} \pair<\eta,\bar \d_{\bar A}\eta> +
     \pair<\bar\eta,\d_A\bar\eta>\nonumber\\
    &&- 2\int_{\Sigma} \pair<\sigma,g^{-1}\bar\psi g> +
     \pair<\bar\sigma,g\psi g^{-1}> - \pair<\sigma,\bar\eta> -
     \pair<\bar\sigma,\eta>~,\label{eq:gcompwzw}
\end{eqnarray}
or, more compactly, by:
\begin{eqnarray*}
\lefteqn{I = I_B [g,A,\bar A] - \int_{\Sigma}
\pair<g(\psi-\eta)g^{-1},\bar \d_{\bar A} g(\psi-\eta)g^{-1}>}\\
& & {} - \int_{\Sigma}
\pair<g^{-1}(\bar\psi-\bar\eta)g,\d_A g^{-1}(\bar\psi-\bar\eta)g>~.
\end{eqnarray*}

\subsection{Witten's action from (\ref{eq:gaugact})}

Let us now start from the gauged SWZW action (\ref{eq:gaugact}) in
superfields, which we recollect here for convenience:
\begin{equation}
I[\tG,\tA,\bar\tA] = I[\tG] - 2 \int_{\Sigma_S} \pair<\tJ,\bar\tA> +
                   \pair<\tA,\bar\tJ> - \pair<\tA,\bar\tA> +
                   \pair<\tA,\tG^{-1}{\bar\tA}\tG>~.\label{eq:gaugactoo}
\end{equation}
In the previous section we have seen how $I[\tG]$ breaks down into
components, yielding $I[g,\psi,\bar\psi,a]$.  The general expressions
of the conserved supercurrents $\tJ$ and $\bar\tJ$ read
\begin{eqnarray*}
\tJ &=& -g\psi g - \theta \left(\d gg^{-1} - g\psi^2 g^{-1}\right) -
      \bar\theta gag^{-1} - \theta\bar\theta \left(\nabla\bar\psi -
      g[a,\psi]g^{-1}\right)~,\\
\bar\tJ &=& g^{-1}\bar\psi g - \theta a + \bar\theta \left(g^{-1}\bar\d
           g - g^{-1}{\bar\psi}^2 g \right) - \theta\bar\theta
           \left(\bar\nabla\psi + [a,g^{-1}\bar\psi g]\right)~.
\end{eqnarray*}
Finally, we parametrise the gauge superfields $\tA$ and $\bar\tA$ as
follows: 
\begin{eqnarray*}
\tA &=& \rho + \theta A + \bar\theta \bar B + \theta\bar\theta
      \lambda~,\\
\bar\tA &=& \bar\rho + \theta B + \bar\theta \bar A + \theta\bar\theta
           \bar\lambda~,
\end{eqnarray*}
where the bosonic components $A$, $\bar A$ correspond to the gauge
fields in the bosonic case, as we will see in a moment.  These new
fields are not all independent, and one can see this in many ways.
One way is to consider the equations of motion of the gauged action,
one of which turns out to be the zero curvature condition for the
gauge superfield.  If we break this equation into components we obtain
several relations between the above gauge field components. Here
though we will follow a different approach.

After some tedious algebra, which includes solving the equation of
motion for the field $a$, the gauged supersymmetric action can be
written as 
\begin{eqnarray}
I[\tG,\tA,\bar\tA] & = &I_B[g,A,\bar A]\nonumber\\
       &&{}+ \int -\pair<\psi,\bcD\psi> - \pair<\bar\psi,
        \cD\bar\psi> + 2\pair<\cD\bar\psi,\bar\rho> +
        2\pair<\rho,\bcD\psi>\nonumber\\
       &&{} + \int 2\pair<\rho + g(\psi-\rho)g^{-1},
        \bar\lambda> + 2\pair<\lambda,\bar\rho -
        g^{-1}(\bar\psi + \bar\rho)g>\nonumber\\
       &&{} + \int \left(-\pair<B+\bar B,B+\bar B>-2\pair<\bar
        B,[g^{-1}(\bar\psi+\bar\rho)g,\rho]>\right.\nonumber\\  
       &&\left. {}+ 2\pair<B,[g(\psi-\rho)g^{-1},\bar\rho]> +
        2\pair<\rho^2,g^{-1}{\bar\rho}^2 g>\right)~,\label{eq:ga1}
\end{eqnarray}
where we have introduced the following covariant derivatives
\begin{eqnarray*}
\cD &\equiv& \ad_g \circ \d_{\cA} \circ \ad_g^{-1} =
\d + [-\d gg^{-1} + g\cA g^{-1},-]~,\\
\bcD &\equiv& \ad_g^{-1} \circ \bar\d_{\bar\cA} \circ \ad_g =
\bar\d + [g^{-1}\bar\d g + g^{-1}\bar\cA g,-]~,
\end{eqnarray*}
where $\cA = A + \rho^2$ and similarly for $\bar\cA$.

The first term in (\ref{eq:ga1}) is nothing but the bosonic gauged WZW
action (I.4.5) with basic fields $g$ and the gauge fields $A$ and
$\bar A$.  In fact, it is easy to see that if we set the fermions to
zero in $I[\tG,\tA,\bar\tA]$ the gauged WZW action reduces to
$I_B[g,A,\bar A]$ --- the bosonic combination $B+\bar B$ also appears
but it is trivially eliminated by its trivial equations of motion.
Hence the $A$, $\bar A$ components of the gauge superfields correspond
indeed to the gauge fields from the bosonic case, at least with
fermions put to zero.

The second line in (\ref{eq:ga1}) can be rearranged to give (up to an
overall sign)
\begin{displaymath}
\pair<(\psi-\rho),\bcD(\psi-\rho)> + \pair<(\bar\psi+
\bar\rho),\cD(\bar\psi+\bar\rho)> - \pair<\rho,\bcD\rho> -
\pair<\bar\rho,\cD\bar\rho>~,
\end{displaymath}
which would look like a difference of two kinds of fermionic terms,
were it not for the $g$-dependent covariant derivatives $\cD$ and
$\bcD$.  This fact can be remedied by suitably rewriting these
terms, together with the gauged bosonic action:
\begin{eqnarray*}
\lefteqn{I_B[g,A,\bar A] + \int \pair<\rho,\bcD\rho> + \pair<\bar\rho,
\cD\bar\rho> =}\\
&& \mbox{} I_B[g,\cA,\bar\cA] + \int \pair<\rho,\bar \d_{\bar\cA}\rho>
+ \pair<\bar\rho, \d_{\cA}\bar\rho> + 2\pair<\rho^2,{\bar\rho}^2> -
2\pair<\rho^2,g^{-1}{\bar\rho}^2 g>~.
\end{eqnarray*}
This gives us back the bosonic action $I_B$, written in terms of
modified bosonic gauge fields $\cA$ and $\bar\cA$, whereas the
resulting fermionic terms describe $\gh$-fermions minimally coupled to
the $\gh$-valued gauge fields $\cA$, $\bar\cA$ through the covariant
derivatives.  The full action can be written as
\begin{eqnarray}
I[\tG,\tA,\bar\tA] &=&I_B[g,\cA,\bar\cA]\nonumber\\
  &&{}+ \int -\pair<(\psi-\rho),\bcD(\psi-\rho)> -
   \pair<(\bar\psi+\bar\rho),  
   \cD(\bar\psi+\bar\rho)>\nonumber\\ 
  &&{}+ \int \pair<\rho,\bar \d_{\bar\cA}\rho> +
   \pair<\bar\rho,\d_{\cA}\bar\rho>\nonumber\\
  &&{}+ \int 2\pair<\rho + g(\psi-\rho)g^{-1},
   \bar\lambda> + 2\pair<\lambda,\bar\rho -
   g^{-1}(\bar\psi + \bar\rho)g>\nonumber\\
  &&{}+ \int - \pair<B+\bar B,B+\bar B> -
   2\pair<\bar B,[g^{-1}(\bar\psi+\bar\rho)g,
   \rho]>\nonumber\\ 
  &&{}+ 2\pair<B,[g(\psi-\rho)g^{-1},\bar\rho]> +
   2\pair<\rho^2,{\bar\rho}^2>~,\label{eq:ga2}
\end{eqnarray}

Notice that the $\lambda$, $\bar\lambda$ gauge fields have no kinetic
terms (dynamics), but rather play the role of Lagrange multipliers,
imposing the following constraint equations on the fermionic fields
\begin{equation}
\bar\rho = g^{-1}(\bar\psi + \bar\rho)g|_{\gh}~, \qquad 
- \rho = g(\psi - \rho)g^{-1}|_{\gh}~,\label{eq:heqmo}
\end{equation}
where $|_{\gh}$ denotes the orthogonal projection $\gg\cong \gh\oplus
\gh^\perp \to\gh$.

If we introduce these equations back in the gauged action and we solve
the equation of motion for the field $B+\bar B$ we obtain
%\begin{displaymath}
%\int - \pair<B+\bar B,B+\bar B> - 2\pair<B+\bar
%                                          B,[\rho,\bar\rho]>~, 
%\end{displaymath}
\begin{displaymath}
B+\bar B=-[\rho,\bar\rho]~,
\end{displaymath}
which will cancel the quartic term in $\rho$, $\bar\rho$.  At this
moment we are left only with the bosonic gauged action and the
fermionic terms, corresponding roughly to minimally coupled $\gg$- and
$\gh$-fermions.  In order to see what the equations of motion
(\ref{eq:heqmo}) impose on the $\gg$-fermions, we will decompose $\gg$
as $\gg = \gh \oplus \gh^\perp$:
\begin{eqnarray*}
\pair<(\psi-\rho),\bcD(\psi - \rho)>_{\gg} &=& 
\pair<g(\psi - \rho)g^{-1}, \bar \d_{\bar\cA} g(\psi -
\rho)g^{-1}>_{\gg}\\
&=& \pair<\rho,\bar \d_{\bar\cA} \rho>_{\gh} + \pair<g(\psi -
\rho)g^{-1}, \bar \d_{\bar\cA} g(\psi - \rho)g^{-1}>_{\gg/\gh}\\
\pair<(\bar\psi + \bar\rho), \cD (\bar\psi + \bar\rho)>_{\gg} &=&
\pair<g^{-1}(\bar\psi + \bar\rho)g, \d_{\cA} g^{-1}(\bar\psi +
\bar\rho)g>_{\gg} \\
&=& \pair<\bar\rho,\d_{\cA} \bar\rho>_{\gh} + \pair<g^{-1}(\bar\psi +
\bar\rho)g,\d_{\cA} g^{-1}(\bar\psi + \bar\rho)g>_{\gg/\gh}~,
\end{eqnarray*}
where we use the suggestive notation $\gg/\gh$ to mean $\gh^\perp$.

The first of the two terms in each RHS clearly cancel the similar
terms in (\ref{eq:ga2}) whereas the remaining terms can be rewritten in
a more compact form if we define new $\gg/\gh$-valued fermionic fields
\begin{displaymath}
\Psi_{\gg/\gh} \equiv g(\psi - \rho)g^{-1}|_{\gg/\gh}\qquad\text{and}
\qquad \bar\Psi_{\gg/\gh} \equiv g^{-1}(\bar\psi +
\bar\rho)g|_{\gg/\gh}~.
\end{displaymath}
Putting all this together, we arrive at the action
\begin{equation}
I[\tG,\tA,\bar\tA] = I_B[g,\cA,\bar\cA] - \int \pair<\Psi_{\gg/\gh},
\bar \d_{\bar\cA}\Psi_{\gg/\gh}> + \pair<\bar\Psi_{\gg/\gh},  
\d_{\cA}\bar\Psi_{\gg/\gh}>~,\label{eq:witac}
\end{equation}
which was introduced by Witten in \cite{Witten}.

Notice nevertheless that the basic fields entering in this action are
not the ones that we would have naively expected.  Indeed, the
$\gg/\gh$-fermions do not coincide with the $\gg/\gh$ subset of the
original $\gg$-fermions, rather they differ by a shift and an axial
gauge transformation from these.  Also, the bosonic gauge fields
$\cA$, $\bar\cA$ differ from the gauge fields that appear in the
bosonic gauged WZW action by a shift quadratic in the fermionic gauge
fields $\rho$, $\bar\rho$.

\subsection{Equivalence with $I[g,\psi,A,\sigma,\rho]$}

We have now two classical gauged WZW actions: the action
(\ref{eq:gcompwzw}) obtained by gauging (\`a la Noether) the component
SWZW action (\ref{eq:compwzw}), and the Witten action (\ref{eq:witac})
obtained by breaking down the superfield gauged SWZW action
(\ref{eq:gaugact}) into components.  The natural question arises whether
they are indeed equivalent, and in this subsection we show how to
relate the two.

In order to do this we go back to the gauged action (\ref{eq:ga2}) and
we make a change of variables, replacing the fields $\lambda$,
$\bar\lambda$ with the following combinations:
\begin{displaymath}
\tau \equiv \lambda + [\bar B,\rho]~,\qquad 
\bar\tau \equiv \bar\lambda - [B,\bar\rho]~.
\end{displaymath}
If we introduce this back in the action, and we solve the algebraic
equation of motion for $B+\bar B$ we obtain:
\begin{eqnarray}
I[\tG,\tA,\bar\tA] &=&I_B[g,\cA,\bar\cA]\nonumber\\
    &&{}- \int \pair<(\psi-\rho),\bar\cD(\psi-\rho)> +
     \pair<(\bar\psi+\bar\rho),  
     \cD(\bar\psi+\bar\rho)>\nonumber\\ 
    &&{}+ \int \pair<\rho,\bar \d_{\bar\cA}\rho> +
     \pair<\bar\rho,\d_{\cA}\bar\rho> + 2 \pair<\tau,\bar\rho> + 2
\pair<\rho,\bar\tau>\nonumber\\
    &&{}- 2 \int \pair<\tau,g^{-1}(\bar\psi + \bar\rho)g> +
\pair<\bar\tau,g(\psi-\rho) g^{-1}>~.\label{eq:gctoo}
\end{eqnarray}
It is easy now to compare this action with the one obtained in the
previous section and show their equivalence.  Indeed, the following
dictionary provides the equivalence:

\begin{center}
\begin{tabular}{||c|c||c|c||c|c||c|c||}
\hline
(\ref{eq:gcompwzw}) & (\ref{eq:gctoo})
&(\ref{eq:gcompwzw}) & (\ref{eq:gctoo})
&(\ref{eq:gcompwzw}) & (\ref{eq:gctoo})
&(\ref{eq:gcompwzw}) & (\ref{eq:gctoo})\\
\hline
$\psi$&$\psi-\rho$&$A$&$\cA$&$\sigma$&$\tau$&$\eta$&$-\rho$\\
$\bar\psi$&$\bar\psi+\bar\rho$&$\bar A$&$\bar\cA$&$\bar\sigma$
&$\bar\tau$&$\bar\eta$&$\bar\rho$\\
\hline
\end{tabular}
\end{center}

In summary, gauging the component SWZW model yields the same theory as
the gauging the superfield SWZW model, and both theories are
equivalent to the one written down by Witten in \cite{Witten}.

\section{Nonreductive Kazama-Suzuki models}

Under certain circumstances the $N{=}1$ coset theory admits an extra
supersymmetry giving rise to an $N{=}2$ coset.  For $\gg$ a reductive
Lie algebra this is the celebrated Kazama-Suzuki construction
\cite{KazamaSuzuki} (see also \cite{Schweigert}).  The purpose of this
section is to extend this construction to the case of self-dual Lie
algebras.

\subsection{$N{=}1$ coset construction in components}

In Section 5 we have studied in considerable detail the expression in
components of the $N{=}1$ affine algebra $\widehat\gg_{N{=}1}$, both
in the coupled (\ref{eq:ktcb1})-(\ref{eq:ktcb3}) and in the decoupled
(\ref{eq:bca})-(\ref{eq:bca3}) basis, and the ones of the two generators
of the $N{=}1$ superconformal algebra (in terms of these currents).
Also, in Section 4 we have considered the $N{=}1$ coset construction,
written in terms of superfields.  In order to proceed further and
investigate the existence of an $N{=}2$ extension to this $N{=}1$
superconformal algebra we need to start with the $N{=}1$ coset theory
written in components.  It is convenient to use a modified decoupled
basis.  If
\begin{displaymath}
\tJ_i (Z) = \Psi_i (z) + \theta I_i (z)~,
\end{displaymath}
then we define a modified bosonic current as follows:
\begin{displaymath}
\widetilde J_i (z) \equiv I_i(z) - \half \Omega^{jl}{f_{ij}}^k
                       (\Psi_k\Psi_l) (z)~.
\end{displaymath}
Notice that this differs from $J_i(z)$ defined in (\ref{eq:deccurr}),
but it is nevertheless decoupled from the $\gh$-fermions:
\begin{displaymath}
\widetilde J_i(z) \Psi_j(w) = \reg~.
\end{displaymath}
These currents define a realisation of an affine Lie algebra
$\widehat\gh$
\begin{displaymath}
\widetilde J_i (z)\widetilde J_j (w) =
\ope[2][\Omega_{ij}-\half\kappa^{\gh}_{ij}] + \ope[1][{f_{ij}}^k
\widetilde J_k (w)] + \reg~,
\end{displaymath}
where the shift in the metric is now proportional to $\kappa^{\gh}$,
the Killing form for $\gh$.

In this basis, the $\gh$ $N{=}1$ Virasoro generators read:
\begin{eqnarray*}
\sT_{\gh}(z) &=& \half\Omega^{ij}(\widetilde J_i \widetilde J_j)(z) + 
              \half\Omega^{ij}(\d\Psi_i\Psi_j)(z)~,\\
\sG_{\gh}(z) &=& \Omega^{ij}(\widetilde J_i \Psi_j)(z) - \fr{1}{6}
f^{ijk} (\Psi_i(\Psi_j\Psi_k))(z)~.
\end{eqnarray*}
The $N{=}1$ coset theory generated by $\sG\equiv \sG_{\gg/\gh} =
\sG_{\gg}-\sG_{\gh}$ and $\sT\equiv \sT_{\gg/\gh} = \sT_{\gg}-\sT_{\gh}$
satisfies the algebra
\begin{eqnarray*}
\sT(z)\sT(w) &=& \ope[4][\half c] + \ope[2][2\sT(w)] + \ope[1][\d \sT(w)]
+ \reg~,\\
\sT(z)\sG(w) &=& \ope[2][\fr{3}{2} \sG(w)] + \ope[1][\d \sG(w)] +
\reg~,\\ 
\sG(z)\sG(w) &=& \ope[3][\fr{2}{3} c] + \ope[1][2\sT(w)] + \reg~,
\end{eqnarray*}
with the central charge $c$ given by (\ref{eq:cosetcc}).

\subsection{$N{=}2$ superconformal cosets}

Now we want to solve the following problem: We want to determine the
conditions under which these $N{=}1$ theories possess an $N{=}2$
superconformal symmetry.  In other words, we want to determine the
conditions under which we can define two new operators, say $\sG^2$
and $\sJ$, such that $(\sT,\sG^+,\sG^-,\sJ)$ satisfy an $N{=}2$
superconformal algebra, where
\begin{displaymath}
\sG^+ = \half (\sG^1 + i \sG^2)~,\qquad 
\sG^- = \half (\sG^1 - i \sG^2)~,
\end{displaymath}
whereas $\sT$ and $\sG^1=\sG$ are the $N{=}1$ coset generators
introduced in the last paragraph. 

In order to do this we will make use of the following characterisation
of the $N{=}2$ Virasoro algebra, proven independently in \cite{FN=2}
and \cite{GetzlerMT}.  The result states that the minimal data
necessary to guarantee the existence of an $N{=}2$ superconformal
algebra consists of two fields $\sG^{\pm}(z)$ satisfying
\begin{eqnarray*}
\sG^{\pm}(z)\sG^{\pm}(w) &=& \reg~,\\
\sG^+(z)\sG^-(w) &=& \ope[3][\fr{1}{3} c] + \ope[2][\sJ(w)] +
\ope[1][\sT(w)+\half\d \sJ(w)] +\reg~,
\end{eqnarray*}
which defines the central charge $c$, and the operators $\sJ$ and $\sT$;
and also such that
\begin{equation}
\sJ(z)\sG^{\pm}(w) = \ope[1][\pm \sG^{\pm}(w)] + \reg~.\label{eq:jgope}
\end{equation}
In other words, provided the above OPEs are satisfied,
$(\sT,\sG^{\pm},\sJ)$ will satisfy an $N{=}2$ superconformal algebra
with central charge $c$.

In our case $\sG^1$ has a rather simple expression. We split $\gg =
\gh \oplus \gh^\perp$, as usual and we introduce bases $\{X_i\}$ and
$\{X_\alpha\}$ for $\gh$ and $\gh^\perp$ respectively.  Of course, as
vector spaces (and even as $\gh$-modules) $\gh^\perp \cong \gg/\gh$,
and we will on occasion allow ourselves to use $\gg/\gh$ as a
shorthand for the subspace $\gh^\perp \in\gg$.  Then $\sG^1$ can be
written in terms of the $\gg/\gh$-fields:
\begin{displaymath}
\sG^1 = \Omega^{\alpha\beta}(J_{\alpha}\Psi_{\beta}) - \fr{1}{6}
f^{\alpha\beta\gamma} (\Psi_{\alpha}(\Psi_{\beta} \Psi_{\gamma}))~.
\end{displaymath}

From the above discussion it follows that we need another fermionic
field $\sG^2(z)$ such that
\begin{eqnarray*}
\sG^1(z)\sG^2(w) &=& \ope[2][2i\sJ(w)] + \ope[1][i\d \sJ(w)] + \reg~,\\
\sG^2(z)\sG^2(w) &=& \ope[3][\fr{3}{2}c] + \ope[1][2\sT(w)] + \reg~,
\end{eqnarray*}
and then, having determined the $U(1)$ current $\sJ(z)$ from the first
OPE, we will have to impose the additional OPEs between $\sJ$ and
$\sG^{1,2}$.  We will find it convenient to rewrite the above two OPEs
using the $\epsilon$-symbol (with $\epsilon_{12} = 1 = -
\epsilon_{21}$):
\begin{equation}
\sG^i(z)\sG^j(w) = \ope[3][\fr{2}{3} c\delta_{ij}] +
\ope[2][2i\sJ(w)\epsilon_{ij}] + \ope[1][2\sT(w)\delta_{ij} +
i\d \sJ(w)\epsilon_{ij}] + \reg~.\label{eq:gij}
\end{equation}

We start with the following Ansatz for $\sG^2$:
\begin{displaymath}
\sG^2(z) = A^{\alpha\beta}(J_{\alpha}\Psi_{\beta})(z) + \fr{1}{6}
         B^{\alpha\beta\gamma}(\Psi_{\alpha}(\Psi_{\beta}
         \Psi_{\gamma}))(z) + C^{\alpha}\d \Psi_{\alpha}(z)~,
\end{displaymath}
where $A$, $B$, $C$ are still to be determined.  By imposing
(\ref{eq:gij}) and (\ref{eq:jgope}) and after some lengthy computations we
obtain a set of necessary and sufficient conditions which after a lot
of effort can be reduced to the following:

\begin{itemize}
\item[(i)]$C^{\alpha}=0$.
\item[(ii)]The matrix $(A^{\alpha\beta})$ defines an $\gh$-invariant
almost complex structure on $\gg/\gh$:
\begin{equation}
A^{\alpha\beta}A^{\gamma\delta}\Omega_{\beta\gamma} = -
\Omega^{\alpha\delta}~,\qquad 
A^{\alpha\beta} = -A^{\beta\alpha}~,\label{eq:comstr} 
\end{equation}
\begin{equation}
A^{\alpha\beta}{f_{\beta}}^{\gamma i} = A^{\gamma\beta}
{f_{\beta}}^{\alpha i}~.\label{eq:invcs}
\end{equation}
One can easily see this by defining a map $A:\gg/\gh \to \gg/\gh$,
with $A\cdot X_{\alpha}\equiv {A^{\beta}}_{\alpha}X_{\beta}$, where
${A^{\beta}}_{\alpha}=A^{\beta\gamma}\Omega_{\gamma\alpha}$; then the
first equation in (\ref{eq:comstr}) is equivalent with $A^2=-\1$. On the
other hand the antisymmetry of $(A^{\alpha\beta})$ tells us that the
complex structure is compatible with the metric on $\gg/\gh$
\begin{displaymath}
\pair<A\cdot X_{\alpha},A\cdot X_{\beta}> =
\pair<X_{\alpha},X_{\beta}>~.
\end{displaymath}
Finally, the relation (\ref{eq:invcs}) states the $\gh$-invariance of
the complex structure on $\gg/\gh$.  Indeed, one can define an action
of $\gh$ on $\gg/\gh$ by understanding $\gg/\gh$ as $\gh^\perp \subset
\gg$ and using the Lie brackets, and (\ref{eq:invcs}) can be expressed
as the fact that $A$ commutes with the above $\gh$ action:
\begin{displaymath}
A\cdot [X_i, X_{\alpha}] = [X_i, A\cdot X_{\alpha}]~.
\end{displaymath}
\item[(iii)]The coefficient of the cubic term in the expression on
$\sG^2$ is given by  
\begin{displaymath}
B^{\mu\nu\rho} = A^{\mu\alpha}A^{\nu\beta}A^{\rho\gamma}
                             f_{\alpha\beta\gamma}~.
\end{displaymath}
\item[(iv)]Finally, the last condition that we obtain
\begin{displaymath}
f^{\mu\nu\rho} = A^{\mu\alpha}A^{\nu\beta}{f_{\alpha\beta}}^{\rho} +
                 A^{\nu\alpha}A^{\rho\beta}{f_{\alpha\beta}}^{\mu} + 
                 A^{\rho\alpha}A^{\mu\beta}{f_{\alpha\beta}}^{\nu}~, 
\end{displaymath}
may seem formidable at first sight, but it is in fact equivalent to
the vanishing of the Nijenhuis tensor associated to the complex 
structure $A$:
\begin{displaymath}
N(X,Y) \equiv [X,Y] - [AX,AY] + A[X,AY] + A[AX,Y]~.
\end{displaymath}
\end{itemize}

One can give an alternative interpretation to the last two conditions.
For this, let us introduce the following projection operators
\begin{displaymath}
(P^{\pm})^\alpha_\beta = \half (\1^\alpha_\beta \pm {1\over
                          i}A^\alpha_\beta)~,
\end{displaymath}
which allows us to split the complexification $\gt =
(\gh^\perp)^\C$ into subspaces $\gt_+$ and $\gt_-$ defined as the
image of the projectors $P^+$ and $P^-$ respectively.  Introducing
bases
\begin{displaymath}
\{X_{\alpha}^{\pm} = (P^{\pm})_\alpha^\beta X_{\beta}\}
\end{displaymath}
for $\gt_\pm$ respectively, we can then show that (\ref{eq:comstr})
implies that $\gt_\pm$ are (maximally) isotropic, whereas {\em
(iii)\/} and {\em (iv)\/} are equivalent to:
\begin{displaymath}
{}[X_{\alpha}^{\pm},X_{\beta}^{\pm}] = \half i
({f_{\alpha\beta}}^{\gamma} \pm i {B_{\alpha\beta}}^{\gamma}) 
                                     X_{\gamma}^{\pm}~. 
\end{displaymath}
This means that $\gt$ admits a decomposition $\gt = \gt_+ \oplus
\gt_-$ into subspaces which close under the Lie brackets:
\begin{displaymath}
{}[\gt_+,\gt_+] \subset \gt_+~,\qquad [\gt_-,\gt_-] \subset \gt_-~,
\end{displaymath}
which re-states the fact that the complex structure in $\gg/\gh$ is
integrable.

Notice that if $\gh=0$, then the condition of $\gh$-invariance would
be trivially satisfied, and the remaining conditions are precisely the
ones in \cite{NouriN=2} (see also \cite{Fc=9N=2}).  In that case,
$\gt= \gg^{\C}$ and $(\gt,\gt_+,\gt_-)$ would be a Manin triple.  In
the more general case, what we have is that $\gt_\pm$ are isotropic
subalgebras of $\gg = \gh\oplus\gt_+\oplus\gt_-$.  In the reductive
case, Getzler \cite{GetzlerMT} has shown that there is indeed an
honest Manin triple underlying the coset construction, with ``double''
given by $\gg\oplus(-\gh)$, where $-\gh$ is $\gh$ with the opposite
metric; although the precise relation between Getzler's Manin triple
and the KS construction will be fully elucidated elsewhere.

In summary, provided $\gg/\gh$ has an $\gh$-invariant metric with a
compatible, integrable, $\gh$-invariant complex structure, the
corresponding $N{=}1$ supersymmetric coset possesses an extended
$N{=}2$ superconformal symmetry.  This $N{=}2$ Virasoro algebra is
generated by $\sT$, $\sG^1$, and in addition the two generators
\begin{equation}
\sG^2 = \Omega^{\alpha\beta}(J_{\alpha}\widetilde\Psi_{\beta})
-\fr{1}{6} f^{\alpha\beta\gamma} (\widetilde\Psi_{\alpha}(\widetilde
\Psi_{\beta}\widetilde\Psi_{\gamma}))~,\label{eq:n=2g1}
\end{equation}
where $\widetilde\Psi_{\alpha}={A^{\beta}}_{\alpha}\Psi_{\beta}$; and
the $U(1)$ current whose expression turns out to be
\begin{equation}
2i\sJ = A^{\alpha\beta}(\Psi_{\alpha}\Psi_{\beta}) - A^{\gamma\delta}
      {f_{\gamma\delta}}^c I_c~,\label{eq:n=2j}
\end{equation}
whereas the central charge is given by (\ref{eq:cosetcc}).  Notice that
it is not possible to assemble $\sJ$ and $\sG^2$ given by
(\ref{eq:n=2g1}) and (\ref{eq:n=2j}) above into a superfield depending
polynomially in the original superfields $\tJ_a$, whence the need to
work in components.

\subsection{The BRST invariance of the $N{=}2$ generators}

We have shown above that the gauged supersymmetric WZW model describes
the $N{=}1$ coset theory.  It thus makes sense that any extended
symmetry of the $N{=}1$ Virasoro algebra which the coset theory
admits, must be already present (maybe up to BRST-exact terms) among
the BRST-invariant fields in the WZW model.  Therefore we expect that
the $N{=}2$ extension, whenever it exists, must be BRST-invariant or,
in this case, since they don't involve the ghosts, actually gauge
invariant.  Since $\sJ$ and $\sG^1$ generate the rest of the $N{=}2$
Virasoro algebra, and $\sG^1$ is already BRST-invariant, all we need
to show is that $\sJ$ is BRST-invariant.

For this we have to first work out the expression of the BRST current
in components.  A convenient parametrisation of the ghost superfields
is given by:
\begin{displaymath}
\tB_i(Z) = \beta_i(z) - \theta b_i(z)\qquad\hbox{and}\qquad
\tC^i(Z) = - c^i(z) - \theta \gamma^i(z)~,
\end{displaymath}
where $(\beta_i,\gamma^i)$ are bosonic fields with weights
$(\half,\half)$ and $(b_i,c^i)$ are fermionic fields with weights
$(1,0)$.  Their OPEs can be read from the ones of $(\tB_i,\tC^i)$:
\begin{displaymath}
b_i(z) c^j(w) = \ope[1][\delta_i^j]\qquad\hbox{and}\qquad
\beta_i(z) \gamma^j(w) = \ope[1][\delta_i^j]~.
\end{displaymath}

The BRST current $q(z)$ is the $\theta$-component of the superfield in
(\ref{eq:jbrst}):
\begin{displaymath}
q(z) = (I_i+\tilde I_i) c^i - (\psi_i+\tilde\psi_i)
\gamma^i + {f_{ij}}^k \beta_k c^i \gamma^j - \half {f_{ij}}^k b_k c^i
c^j~,
\end{displaymath}
and BRST transformations on fields are given by $d\phi = [q,\phi]_1$
in the notation of the Appendix.

It is now a simple matter to prove that the expression (\ref{eq:n=2j})
for $\sJ(z)$ is BRST invariant.  In fact, this follows trivially from
the $\gh$-invariance of the complex structure.  To this effect, notice
that the BRST transformation of the coset fermions $\psi_\alpha$ is
precisely an $\gh$-gauge rotation:
\begin{displaymath}
d\psi_\alpha = - {f_{i\alpha}}^\beta \psi_\beta c^i~.
\end{displaymath}
Hence any $\gh$-invariant tensor contracted with coset fermions is
automatically BRST-invariant.  This, together with the identity
\begin{displaymath}
A^{\alpha\beta} {f_{\alpha\beta}}^c {f_{c i}}^b = 0~,
\end{displaymath}
is enough to show that $\sJ$ is BRST-invariant.  This proves that the
gauged supersymmetric WZW model does provide a lagrangian realisation of
the $N{=}2$ coset construction.

One might wonder whether in the same way that the $N{=}1$ coset theory
is induced from a natural $N{=}1$ SCFT involving all three sectors in
the gauged supersymmetric WZW model: $(\sG_{\mathrm{tot}},
\sT_{\mathrm{tot}})$, the same is true for the $N{=}2$ coset.  In
other words, {\em is there a natural BRST invariant $N{=}2$ SCFT
involving the ghosts, extending $(\sG_{\mathrm{tot}},
\sT_{\mathrm{tot}})$, and which is BRST-cohomologous to the one
generated by $(\sJ,\sG^1,\sG^2,\sT)$?}

Let us try to answer this question.  Notice first of all that the
ghost $N{=}1$ Virasoro algebra does extend to an $N{=}2$ with
generators:
\begin{displaymath}
\sJ_{\mathrm{gh}} \equiv \beta_i \gamma^i \qquad\hbox{and}\qquad 
\sG^2_{\mathrm{gh}} \equiv -i \left( b_i\gamma^i - \beta_i\d
c^i\right)~,
\end{displaymath}
in addition to $\sG^1_{\mathrm{gh}} \equiv \sG_{\mathrm{gh}}$ and
$\sT_{\mathrm{gh}}$.  We would therefore need to find a BRST-exact
$\sJ' = \sJ_{\mathrm{gh}} + \cdots$.  It turns out that there is a
unique such $\sJ'$:
\begin{displaymath}
\sJ' \equiv \beta_i \gamma^i + \Omega^{ij} \psi_i\tilde\psi_j =
d\left(-\half \Omega^{ij} \beta_i\psi^-_j\right)~.
\end{displaymath}
Let us define $\sJ_{\mathrm{tot}} \equiv \sJ + \sJ'$.  Similarly let
$\sG^1_{\mathrm{tot}} \equiv \sG_{\mathrm{tot}}$.  It turns out that
these generators do not form generally satisfy an $N{=}2$
superconformal algebra, but they do when $\gh$ is abelian!  The
passage from the total $N{=}2$ superconformal algebra to the coset one
can be understood as the conformal field theoretical manifestation of
the Poisson reduction of a Poisson Lie group.  Details will appear
elsewhere.

\begin{ack}
We are grateful to Roya Moyahee, Henric Rhedin, Kris Thielemans and
George Thompson for useful conversations about this topic.  SS would
like to thank the members of the String Theory group of Queen Mary and
Westfield College and Chris Hull in particular for the hospitality
during part of the time it took to do this work.  JMF would like to
thank the Institute for Theoretical Physics at Stony Brook and
especially Martin Ro\v cek, and the Mathematics Department of Boston
University and especially Takashi Kimura, for their hospitality during
the final stages of this work.

This paper is archived in {\tt hep-th/9605111}.
\end{ack}

\appendix

\section{Supersymmetric OPE Technology}

In this appendix we collect some formulas which are useful in the
computations concerning superspace operator product expansions in
meromorphic superconformal field theory.  See \cite{Kris} for the
computer implementation of these formulas (and their $N{=}2$
extension).

Our superfields are functions $\Phi(Z)$ in a $(1|1)$-superspace whose
points are denoted by $Z = (z,\theta)$, with $z$ even and $\theta$
odd.  The supercovariant derivative is given by $D = {\d\over\d\theta}
+ \theta \d$, where we use the abbreviation $\d$ to mean $\d\over\d
z$. The supercovariant derivative obeys $D^2 = \d$.

Given two points $Z_i = (z_i,\theta_i)$ for $i=1,2$, we define even
and odd superintervals:
\begin{displaymath}
Z_{12} = \hbox{``$Z_1 - Z_2$''} \equiv z_1 - z_2 - \theta_1\theta_2
 \qquad\hbox{and}\qquad \theta_{12} = \hbox{``$Z_{12}^{\shalf}$''}
\equiv \theta_1 - \theta_2~,
\end{displaymath}
where the notation $Z_{12}^{\shalf}$ has been introduced for
convenience.  More generally, $Z_{12}^{n + \shalf} = Z_{12}^n
\theta_{12}$ for any $n\in\Z$.

There exists a supersymmetric analogue of the Cauchy residue calculus.
Defining the measure $dZ \equiv {dz\over 2\pi i} d\theta$, and the
contour integral $\oint_Z$ to refer both to the contour integral about
$z$ and the Berezin integral, Laurent expansions take the form
\begin{displaymath}
\Phi(Z_1) = \sum_{2r\in\Z} Z_{12}^{-r-\shalf}\, \Phi^{(r)}(Z_2)~,
\end{displaymath}
where
\begin{displaymath}
\Phi^{(r)}(Z_2) = \oint_{Z_2} dZ_1\, Z_{12}^r\, \Phi(Z_1)~.
\end{displaymath}
And in particular, Taylor expansions are of the form
\begin{displaymath}
\Phi(Z_1)  = \sum_{2r\in|\Z|} {1\over \lfloor r\rfloor!}\,
Z_{12}^r \, D^{2r}\Phi(Z_2)~,
\end{displaymath}
where
\begin{displaymath}
\oint_{Z_2} dZ_1\, Z_{12}^{-(r+\shalf)}\, \Phi(Z_1) = {1\over \lfloor r
\rfloor!}\, D^{2r}\Phi(Z_2)\qquad\hbox{for $2r\in|\Z|$}~,
\end{displaymath}
where $\lfloor r\rfloor$ denotes the greatest integer $\leq r$.

The superspace operator product expansion takes the form
\begin{displaymath}
\tA(Z_1) \tB(Z_2) = \sum_{2r \in\Z} Z_{12}^{-r}\,
{}[\![\tA,\tB]\!]_r(Z_2)~,
\end{displaymath}
where by definition,
\begin{displaymath}
{}[\![\tA,\tB]\!]_r(Z_2) = \oint_{Z_2} dZ_1 \, Z_{12}^{r-\shalf}\,
\tA(Z_1) \tB(Z_2)~.
\end{displaymath}

If $\tA(Z) = \phi_A(z) + \theta \psi_A(z)$ and $\tB(Z) = \phi_B(z) +
\theta \psi_B(z)$, the brackets $[\![\tA,\tB]\!]_r$ can be written in
terms of the similar brackets of the component fields as follows:
\begin{eqnarray*}
{}[\![\tA,\tB]\!]_n(Z) &=& [\phi_A,\phi_B]_n(z) + \theta \left(
{}[\psi_A,\phi_B]_n(z) + (-)^{|\tA|} [\phi_A,\psi_B]_n(z) \right)\\
{}[\![\tA,\tB]\!]_{n+\shalf}(Z) &=& [\psi_A,\phi_B]_{n+1}(z) - \theta
\left( n [\phi_A,\phi_B]_n(z) + (-)^{|\tA|} [\psi_A,\psi_B]_{n+1}(z)
\right)
\end{eqnarray*}
for every $n\in\Z$, where the brackets $[-,-]_n$ are defined as
usual by the ordinary operator product expansion:
\begin{displaymath}
A(z) B(w) = \sum_n {[A,B]_n(w)\over (z-w)^n}~.
\end{displaymath}

Then either from the identities obeyed by the $[-,-]_n$ or working
directly from the properties of the superspace operator product
expansion, one can derive a set of ``axioms'' obeyed by the
$[\![-,-]\!]_r$.  These axioms encode the properties of identity,
commutativity and associativity of the operator product expansion, as
well as the properties of the normal ordered product and of the
supercovariant derivative $D$.  Since these two operations generate
the operator algebra of the superconformal field theory starting from
a set of generating fields, the above axioms allow us to compute {\em
all\/} $[\![-,-]\!]_r$ knowing only the $[\![-,-]\!]_{r>0}$ of the
generating fields.

It is convenient in what follows to separately discuss the brackets
$[\![-,-]\!]_n$ and $[\![-,-]\!]_{n+\shalf}$, where $n\in\Z$.  We
first have the identity axiom:
\begin{displaymath}
{}[\![\1,\tA]\!]_r = \left\{ \begin{array}{ll}
                             \tA~, & \mbox{ for $r=0$}\\
                               0~, & \mbox{ otherwise}
			     \end{array} \right. ~,
\end{displaymath}
where $\1$ is the identity.

We then have the action of the supercovariant derivative:
\begin{eqnarray*}
&&[\![D\tA,\tB]\!]_n = [\![\tA,\tB]\!]_{n-\shalf}~,\\
&&[\![D\tA,\tB]\!]_{n+\shalf} = -n [\![\tA,\tB]\!]_n~,\\
&&[\![\tA,D\tB]\!]_n = (-)^{|\tA|}\left( D[\![\tA,\tB]\!]_n -
{}[\![\tA,\tB]\!]_{n-\shalf} \right)~,\\
&&[\![\tA,D\tB]\!]_{n+\shalf} = -(-)^{|\tA|}\left(
D[\![\tA,\tB]\!]_{n+\shalf} + n [\![\tA,\tB]\!]_n\right)~.
\end{eqnarray*}
Iterating these relations we find the ones for $\d$:
\begin{eqnarray*}
&&[\![\d\tA,\tB]\!]_n = (1-n) [\![\tA,\tB]\!]_{n-1}~,\\
&&[\![\d\tA,\tB]\!]_{n+\shalf} = -n [\![\tA,\tB]\!]_{n-\shalf}~,\\
&&[\![\tA,\d\tB]\!]_n = \d[\![\tA,\tB]\!]_n + (n-1)
{}[\![\tA,\tB]\!]_{n-1}~,\\
&&[\![\tA,\d\tB]\!]_{n+\shalf} = \d[\![\tA,\tB]\!]_{n+\shalf} + n
{}[\![\tA,\tB]\!]_{n-\shalf}~.
\end{eqnarray*}

Notice that the above relations imply that $D$ (resp. $\d$) is an odd
(resp. even) derivation over all of the $[\![-,-]\!]_r$, for
$2r\in\Z$; that is,
\begin{eqnarray*}
D[\![\tA,\tB]\!]_r &= [\![D\tA,\tB]\!]_r + (-)^{|\tA|+2r}
{}[\![\tA,D\tB]\!]_r\\
\d[\![\tA,\tB]\!]_r &= [\![\d\tA,\tB]\!]_r + [\![\tA,\d\tB]\!]_r~.
\end{eqnarray*}
The sign in the first equation can be understood if we notice that the
$[\![-,-]\!]_r$ brackets for $r - \shalf\in\Z$ have odd parity, so
that the operation $\ad_r\tA \equiv [\![\tA,-]\!]_r$ has parity
$|\ad_r\tA| = |\tA| + 2r$.

Next the commutativity axiom takes the form (for $n\in\Z$):
\begin{eqnarray*}
{}[\![\tA,\tB]\!]_n &=& (-)^{|\tA||\tB| + n} \sum_{m\geq 0}
{(-)^m\over m!} \d^m[\![\tB,\tA]\!]_{n+m}\\
{}[\![\tA,\tB]\!]_{n+\shalf} &=& (-)^{|\tA||\tB| + n} \sum_{m\geq 0}
{(-)^m\over m!} \d^m \left( [\![\tB,\tA]\!]_{n+m+\shalf} -
D[\![\tB,\tA]\!]_{n+m+1} \right)~.
\end{eqnarray*}
In particular, the normal ordered product obeys:
\begin{displaymath}
(\tA\tB) \equiv [\![\tA,\tB]\!]_0 = (-)^{|\tA||\tB|}  \sum_{m\geq 0}
{(-)^m\over m!} \d^m[\![\tB,\tA]\!]_m~,
\end{displaymath}
so that the normal ordered commutator is given by
\begin{displaymath}
(\tA\tB) - (-)^{|\tA||\tB|} (\tB\tA) =  (-)^{|\tA||\tB|} \sum_{m\geq
1} {(-)^m\over m!} \d^m[\![\tB,\tA]\!]_m~,
\end{displaymath}
which coincides with the analogous formula in the nonsupersymmetric
case.

Now we come to the associativity axioms.  In the formulas which
follow, $n,m\in\Z$ and in addition we take $m> 0$ in the first and
third and $m\geq 0$ in the remaining two:
\begin{eqnarray*}
{}[\![\tA,[\![\tB,\tC]\!]_n]\!]_m &=& (-)^{|\tA||\tB|}
[\![\tB,[\![\tA,\tC]\!]_m]\!]_n\\
&&\mbox{} +  \sum_{q\geq 0} {m-1\choose q}
[\![[\![\tA,\tB]\!]_{q+1},\tC]\!]_{m+n-q-1}~,\\
{}[\![\tA,[\![\tB,\tC]\!]_n]\!]_{m+\shalf} &=&
(-)^{|\tB|(|\tA|+1)} [\![\tB,[\![\tA,\tC]\!]_{m+\shalf}]\!]_n\\
&& \mbox{} + \sum_{q\geq 0} {m\choose q}
[\![[\![\tA,\tB]\!]_{q+\shalf},\tC]\!]_{m+n-q}~,\\
{}[\![\tA,[\![\tB,\tC]\!]_{n+\shalf}]\!]_m &=&
(-)^{|\tA|(|\tB|+1)} [\![\tB,[\![\tA,\tC]\!]_m]\!]_{n+\shalf}\\
&& \mbox{} + \sum_{q\geq 0} {m-1\choose q}
\left([\![[\![\tA,\tB]\!]_{q+1},\tC]\!]_{m+n-q-\shalf}\right.\\
&& \qquad \mbox{}  -
\left. [\![[\![\tA,\tB]\!]_{q+\shalf},\tC]\!]_{m+n-q}\right)~,\\
{}[\![\tA,[\![\tB,\tC]\!]_{n+\shalf}]\!]_{m+\shalf} &=&
(-)^{(|\tA|+1)(|\tB|+1)}
[\![\tB,[\![\tA,\tC]\!]_{m+\shalf}]\!]_{n+\shalf}\\
&& \mbox{} + (-)^{|\tA|} \sum_{q\geq 0} {m\choose q}
[\![[\![\tA,\tB]\!]_{q+\shalf},\tC]\!]_{m+n-q+\shalf}\\
&& \mbox{} - (-)^{|\tA|} \sum_{q\geq 0} m {m-1\choose q}
[\![[\![\tA,\tB]\!]_{q+1},\tC]\!]_{m+n-q}~.
\end{eqnarray*}

In particular, these axioms imply that the operation
$[\![\tA,-]\!]_{\shalf}$ is a derivation of parity $|\tA| + 1$ over
all the $[\![-,-]\!]_r$:
\begin{displaymath}
{}[\![\tA,[\![\tB,\tC]\!]_r]\!]_{\shalf} = (-)^{2r|\tA|}
{}[\![[\![\tA,\tB]\!]_{\shalf},\tC]\!]_r + (-)^{(|\tA|+1)(|\tB|+2r)}
{}[\![\tB,[\![\tA,\tC]\!]_{\shalf}]\!]_r~.
\end{displaymath}
Moreover it follows that $[\![\tA,-]\!]_1$ is a derivation of parity
$|\tA|$ over the integral $[\![-,-]\!]_n$, but not over the
half-integral  $[\![-,-]\!]_{n+\shalf}$.  In particular, it is a
derivation over the normal ordered product.

Finally, we arrive at the rearrangement lemma which are crucial in
bringing normal ordered products to a standard form:
\begin{displaymath}
(\tA(\tB\tC)) - (-)^{|\tA||\tB|} (\tB(\tA\tC)) = ((\tA\tB)\tC) -
(-)^{|\tA||\tB|} ((\tB\tA)\tC)~.
\end{displaymath}

\end{document}